# Computational Study of Atomic Mobility in HCP Mg-Al-Zn Ternary Alloys

Jingya Wang[1], Na Li[1], Chuanyun Wang[1], Juan Ignacio Beltran[1],

Javier LLorca[1,2], Yuwen Cui[1,3*]

[1]IMDEA Materials Institute. C/ Eric Kandel 2, 28906 – Getafe, Madrid, Spain.

[2]Department of Materials Science. Polytechnic University of Madrid. 28040 – Madrid, Spain

[3]Institute for Advanced Metallic Materials & School of Materials Science and Engineering, Nanjing Tech University, Nanjing 210009, China

**Abstract**

The experimental data in the literature for the hcp phase of the Mg-Al-Zn ternary system have been critically reviewed. Based on the concentration profiles from the literature, the diffusion coefficients have been re-extracted using the Hall method for the impurity diffusion, and the Sauer-Freise and the Whittle-Green strategies for interdiffusion coefficients in binary and ternary systems, respectively. Moreover, extra interdiffusion coefficients were obtained from the "Darken-type" couples, which present relative maxima or/and minima at the concentration profiles. This information was assessed to obtain an atomic mobility database, by means of DICTRA software in conjunction with the CALPHAD thermodynamic description that is able to reproduce the diffusion couple experiments. Comprehensive comparisons between the calculated results and experimental values show an excellent agreement not only for the diffusion coefficient data, but also for the concentration profiles and the diffusion paths.

**Keyword:** hcp Mg-Al-Zn ternary system, "Darken-type" couple, atomic mobility, DICTRA

## 1. Introduction

Mg-Al, Mg-Zn and Mg-Al-Zn alloys stand among the most popular Mg alloys for structural applications. Both Al and Zn have good solubility in Mg, and form intermetallic precipitates ($Mg_{17}Al_{12}$, $MgZn$ and $Al_5Mg_{11}Zn_4$), leading to an optimum combination of strength and ductility, through solid solution strengthening and precipitation hardening [1]. Moreover, they present very good castability, satisfactory corrosion resistance and are cheaper than other Mg alloys because Al and Zn are very abundant elements [1]. Thus, there is a strong interest to further improve the microstructure and properties of these alloys and virtual processing [2][4] and virtual testing strategies [5] are starting to be used to this end. In the particular case of phase-field modeling of solidification, reliable thermodynamic data as well as kinetic information are critical to achieve good results.

A sound thermodynamic description of the ternary Mg-Al-Zn was provided via CALPHAD (CALculation of Phase Diagram) by Liang et al. [6], taking into account experimental data together with constitutional, thermodynamic and crystallographic literature information. Phase diagrams and properties of different phases can be predicted with this methodology, which has been used to predict diffusion coefficients and to simulate diffusion phenomena in combination with DICTRA (DIffusion Controlled TRAnsformation) [7][9].

However, experimental diffusion data are limited and there is not yet any available atomic mobility database for the hcp phase of Mg-Al-Zn ternary system because of two reasons. The first one comes from the experimental difficulties associated with melting high purity Mg alloys due its high vapor pressure. The second one is induced by the anisotropy of the hcp structure which complicates the diffusion behavior, as compared with cubic structures. Hence, it is important to establish a set of atomic mobility parameters that dictate the kinetic behavior of ternary Mg-Al-Zn alloys and this is the main objective of this investigation. To this end, the experimental diffusion data in the Mg-Al, Mg-Zn and Mg-Al-Zn systems have been critically reviewed, including the impurity diffusion coefficients of Al and Zn in different Mg-based alloy systems along with the interdiffusion coefficients [10][18]. Based on the concentration profiles from the literature, the diffusion coefficients have been re-extracted using the Hall method for impurity diffusion, and the Sauer-Freise and the Whittle-Green strategies for interdiffusion coefficients in binary and ternary systems, respectively. From this information, the atomic mobility for the hcp phase of the Mg-Al, Mg-Zn and Mg-Al-Zn systems was obtained by using the parrot module of the DICTRA software and the diffusion profiles and diffusion paths were simulated from the optimized mobility parameters to provide further insight into the diffusion phenomena in the Mg-Al-Zn system.

## 2. Atomic mobility modeling

Andersson and Agren [8] suggested that the atomic mobility $M_i$ could be expressed as a function of the absolute temperature $T$ as:

$$M_i = M_i^0 \exp\left(\frac{-Q_i}{RT}\right)\frac{1}{RT} = \exp\left(\frac{RT\ln M_i^0}{RT}\right)\exp\left(\frac{-Q_i}{RT}\right)\frac{1}{RT} \tag{1}$$

where $M_i^0$ is the frequency factor with $i$ standing for the species, $Q_i$ the activation energy and R the gas constant. Note that the mobility parameters, $-Q_i$ and $RT\ln M_i^0$ can be grouped into one single parameter, $\Psi_i = -Q_i + RT\ln M_i^0$, in the DICTRA notation [19]. Following the phenomenological CALPHAD approach, the parameter $\Psi_i$ is assumed to depend on the composition through a Redlich-Kister polynomial [20][21]:

$$\Psi_i = \sum_p x_p \Psi_i^p + \sum_p \sum_{q>p} x_p x_q \left[\sum_{r=0,1,2...} {}^r\Psi_i^{p,q}\left(x_p - x_q\right)^r\right] + \sum_p \sum_{q>p} \sum_{v>q} x_p x_q x_v \left[v_{pqv}^s \, {}^s\Psi_i^{p,q,v}\right] \quad s=i,j,k \tag{2}$$

where $x_p$ is the mole fraction of species $p$, and $\Psi_i^p$ is the value of $\Psi_i$ for the pure species $i$. ${}^r\Psi_i^{p,q}$ and ${}^s\Psi_i^{p,q,v}$ stand for the binary and ternary interaction parameters, while the parameter $v_{p,q,v}^s$ is given by $v_{p,q,v}^s = x_s + (1 - x_p - x_q - x_v)/3$. For a binary system, only the first two terms of the right-hand side of equation (2) have to be considered.

The tracer diffusion coefficient $D_i^*$ is rigorously related to the atomic mobility by the simple relation, $D_i^* = RTM_i$ where $M_i$ is atomic mobility of species $i$. In addition, if chemical diffusion takes place in presence of compositional gradients, their effect can be derived by introducing an accurate thermodynamic description of the material system according to:

$$D_{i,j}^n = \sum_{i=1}^{n-1}(\delta_{ik} - x_k)x_i M_i\left(\frac{\partial \mu_i}{\partial x_j} - \frac{\partial \mu_i}{\partial x_k}\right) \tag{3}$$

where $D_{i,j}^n$ are the interdiffusion coefficients in the alloys, $\delta_{ik}$ the Kronecker delta (=1 when i=k, and 0 otherwise) and $\mu_i$ the chemical potential of species $i$. The superindex $n$ stands for the dependent species throughout the work. In this work, the parameters were obtained by fitting to the experimental diffusion coefficients available in the literature using an optimization algorithm.

## 3. Evaluation of experimental diffusivities

### 3.1 Impurity diffusion

The experimental impurity diffusion coefficient of Al into polycrystalline hcp-Mg has been determined by different authors. Brennan [11][12] measured the Al impurity diffusion in polycrystalline hcp-Mg via depth profiling with secondary ion mass spectrometry in the temperature range 573K-673K. The thin film method and thin film solution of the diffusion equation were used to determine the diffusion coefficients. However, some contributions from its small grain size was introduced during the measurement process and these data were considered to be unreliable as detailed in their research. Afterwards, in order to improve the experimental results, Brennan [13] carried out experiments in polycrystalline hcp-Mg using solid-to-solid diffusion couples at 573K, 623K and 673K. The impurity diffusion coefficients were re-calculated according to the Boltzmann-Matano method based on the concentration profiles measured by electron probe micro analysis (EPMA). However, the solubility of Al in hcp-Mg was so small at 573K that the accuracy of these measurements was poor, and the experimental values at 573K were not used in the current work.

Kammerer et.al. [14][15] studied the diffusion behavior of Al and Zn in Mg solid solution using solid-to-solid diffusion couples in polycrystalline hcp-Mg in the temperature range from 623K-723K. The impurity diffusion coefficients were also calculated from the concentration profiles, and they were used in our optimization. In addition, the impurity diffusion coefficients were also obtained using the Hall method in the present work from these experimental concentration profiles as listed in Table 1.

Table 1. Al and Zn impurity diffusion coefficients re-calculated in this work using the Hall method from the experimental data [15]

| Temperature(K) | $D_{Al}^{Mg}(m^2/s)$ | $D_{Zn}^{Mg}(m^2/s)$ |
|---|---|---|
| 623K | $1.81*10^{-16}$ | $3.31*10^{-15}$ |
| 673K | $0.92*10^{-15}$ | $2.21*10^{-14}$ |
| 723K | $5.12*10^{-15}$ | $5.79*10^{-14}$ |
| $D_0(m^2/s)$ | $1.38*10^{-4}$ | $8.68*10^{-4}$ |
| Q(kJ/mol) | 143.02 | 138.74 |

Regarding the impurity diffusion of Zn in hcp-Mg, there are four main experimental studies. Lal [16] and Čermák [17] diffused radioactive $^{65}Zn$ isotope into polycrystalline hcp-Mg in the temperature range 498K-848K. The concentration depth profile was measured via serial sectioning and residual activity methods. Kammerer et al. [14][15] analyzed the diffusion couples between polycrystalline hcp-Mg and Mg-6.at% Zn alloy in the temperature range 623K-723K. In addition, the impurity diffusion coefficient was re-calculated using the Hall method from the experimental data of Kammerer et al., and they are depicted in Table 1.

### 3.2 Interdiffusion coefficient

In the Mg-Zn binary system, the solubility of Zn in hcp-Mg is quite narrow as well as in the Mg-Al binary system. Thus, there are few experimental studies of interdiffusion in this system. The interdiffusion in Mg binary solid solutions, Mg(Al) and Mg(Zn) was investigated by Kammerer et al. [14][15] in the temperature 623K-723K by means of solid-solid diffusion couples. The interdiffusion coefficients were determined via the Boltzmann-Matano method. All these experimental results were used in our optimization, especially the original concentration profiles that play a significant role in the optimization process.

Experimental results of the diffusion of Al and Zn in the hcp-solid solution phase of Mg-Al-Zn ternary system were carried out by Kammerer et al. [18] in the temperature range 673K-723K, using different solid-to-solid diffusion couples, (I: Mg-9at.%Al/Mg-3at%.Zn, II: Mg-3at.%Al/Mg-1at.%Zn, III: Mg/Mg-3at.%Al-0.5at.%Zn and IV: Mg/Mg-1at.%Al-1at.%Zn). Only two sets of interdiffusion coefficients were obtained at the intersection compositions of couple II and couple III together with couple II and couple IV at 723K, as well as one set of couple II and couple III at 673K. The main-interdiffusion coefficients $D_{Al,Al}^{Mg}$ and $D_{Zn,Zn}^{Mg}$ were positive, and the cross interdiffusion coefficients $D_{Al,Zn}^{Mg}$ and $D_{Zn,Al}^{Mg}$ were negative. Besides, the main-interdiffusion coefficients $D_{Zn,Zn}^{Mg}$ were determined based on the couple I, II and III at both temperatures at the terminal ends of the concentration profiles where the relative maxima and minima were found, as in the "Darken-type" couple described in the Appendix. Except for this, the impurity diffusion coefficients of Al in the Mg(Zn) solid solution and Zn in the Mg(Al) solid solution have also been confirmed from the concentration profiles. All of these data were utilized in the assessment procedure to obtain a set of self-consistent parameters that reproduces the experimental results.

In addition, the raw experimental EPMA profiles measured by Kammerer et al. [15],[18] in the Mg-Al, Mg-Zn and Mg-Al-Zn systems were used for the optimization procedure. They were firstly represented by a mathematical superposition of error functions [22], denominated Error Function Expansion (ERFEX), according to:

$$X(r) = \sum_i a_i erf[b_i r + c_i] \qquad (4)$$

where $a$, $b$ and c are the fitting parameters. The error function expansion provides a very close reproduction of EPMA profiles while eliminating point-to-point concentration fluctuations, yet allows more sound physical meaning by applying the error function to diffusion, and this is important to obtain very accurate results. From these smoothed concentration profiles, the diffusion coefficients were extracted using the Sauer-Freise method and the Whittle-Green method at the intersection common compositions and the relative maxima and/or minima.

## 4. Atomic Mobility Assessment

A sound thermodynamic description of the Mg-Al-Zn system was well developed by Liang.et al. [6]. The thermodynamic parameters represented accurately the experimental phase diagram and thermodynamic properties. Therefore, these thermodynamic parameters have been directly adopted in our work. The atomic mobility parameters were assessed on the basis of the available experimental data selected in our work via Parrot module of the DICTRA software package, and the optimized parameters have been summarized in Table 2.

### 4.1 The Mg-Al binary system

Bryan et al. [10] performed a complete study of the atomic mobility in hcp Mg-Al alloys based on critical revision of the experimental data. However, their mobility parameters could not reproduce the experimental results very well, especially the concentration profiles. Hence, the mobility parameters were reassessed in this work on the basis of the interdiffusion coefficients and impurity diffusion coefficients calculated by our own code and the raw EPMA experimental concentration profiles in order to improve the accuracy of the results.

The mobility parameter corresponding to self-diffusion in hcp-Mg was evaluated by Bryan et al. [10], and satisfied the majority of diffusional experimental data. Thus it has been utilized in the present work. Regarding the end-member for the Mg impurity diffusion in hcp Al ($\Psi_{Mg}^{Al}$), it cannot be determined from experimental data, because there is no stable hcp structure for Al. Thus, it was directly taken from Bryan et al. [10]. The end-member parameter $\Psi_{Al}^{Mg}$ was optimized from the experimental data of the impurity diffusion coefficient [11]-[15], including results calculated in the present work using the Hall method from the concentration profiles from [15], which were firstly smoothed by the ERFEX method. And these data were given a higher weight during the optimization procedure due to their accuracy. The binary interaction parameters, $^0\Psi_{Al}^{Al,Mg}$ and $^1\Psi_{Al}^{Al,Mg}$ were assessed from the experimental interdiffusion coefficients and the raw concentration profiles.

**Table 2. Optimized atomic mobility parameters obtained in this work**

| Mobility | Parameter (J/mol) | Reference |
|---|---|---|
| *Mobility of Mg* | | |
| $\Psi_{Mg}^{Mg}$ | -125077 - 88.17 * T | [10] |
| $\Psi_{Mg}^{Al}$ | -941760 + 63.18 * T | [10] |
| $\Psi_{Mg}^{Zn}$ | -91760 - 93.60 * T | This work |
| *Mobility of Al* | | |
| $\Psi_{Al}^{Al}$ | -73360 – 95.08 * T | [26] |
| $\Psi_{Al}^{Zn}$ | -86960 - 93.60 * T | This work |
| $\Psi_{Al}^{Mg}$ | -143015 -73.88 * T | This work |
| $^0\Psi_{Al}^{Al,Mg}$ | 581278 +114.95 * T | This work |
| $^1\Psi_{Al}^{Al,Mg}$ | 820002 - 238.59 * T | This work |
| $^0\Psi_{Al}^{Zn,Mg}$ | -3918294 + 6114.65*T | This work |
| *Mobility of Zn* | | |
| $\Psi_{Zn}^{Zn}$ | -91760 - 93.60 * T | [26] |
| $\Psi_{Zn}^{Al}$ | -91760 - 93.60 * T | [26] |
| $\Psi_{Zn}^{Mg}$ | -138743 - 58.61*T | This work |
| $^0\Psi_{Zn}^{Zn,Mg}$ | 5183886 + 866.63 * T | This work |
| $^1\Psi_{Zn}^{Zn,Mg}$ | -6050970 + 324.68 * T | This work |
| $^0\Psi_{Zn}^{Al,Mg}$ | 969245 + 369.11 * T | This work |
| $^1\Psi_{Zn}^{Al,Mg}$ | 1616532 - 241.02 * T | This work |

The predictions of the impurity diffusion of Al in hcp-Mg obtained from the optimized atomic mobility parameters are plotted in Fig. 1, together with the corresponding experimental data. They can be expressed as an Arrhenius-type equation, which depends on the frequency factor $D_0$ and the activation energy $Q$, which were given by $1.38*10^{-4}$ m$^2$/s and 143.02 kJ/mol, respectively. The predictions obtained from the optimized set of mobility parameters are in good agreement with the experimental data obtained from the diffusion couple [15].

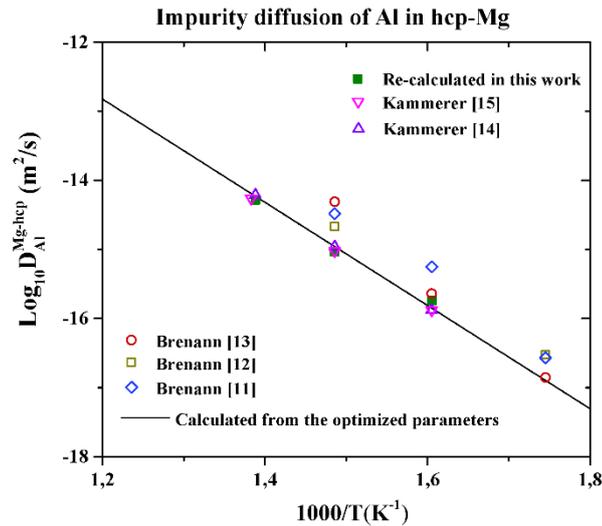

Figure 1. Variation of the Al impurity diffusion coefficient in pure Mg ($D_{Al}^{Mg}$) as a function of the inverse of the absolute temperature. The solid line stands for the predictions with the optimized parameters, which is compared with the experimental data [11]-[15] as well as with the data re-calculated in this work from the concentration profiles.

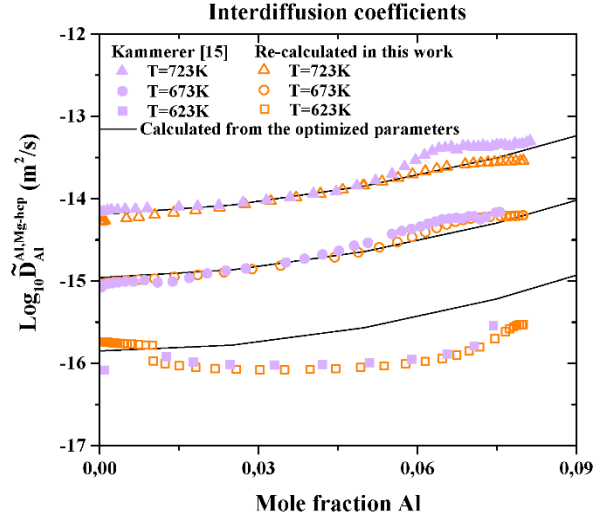

Figure 2 Interdiffusion coefficients of the hcp Mg-Al couple ($D_{Al}^{Al,Mg}$) as a function of the Al content at 623K, 673K and 723K. The solid lines stand for the predictions with the optimized parameters which are compared with the experimental values [15] as well as with the data re-calculated in this work from the concentration profiles.

The interdiffusion coefficients of the hcp Mg-Al in the temperature range 623K-723K obtained from the optimized mobility parameters are plotted in Fig. 2 as a function of the Al content together with the experimental values from [15]. They are in good agreement at high temperature, but the calculated values overestimate the experimental results at 623K because the experimental values of the interdiffusion coefficient remain practically constant with the Al content. This experimental result contradicts the empirical theory that establishes that the diffusion coefficients should increase if the addition of alloying elements decreases the alloy melting point. Thus, the experimental data at 623K are not very reliable and they were given a lower weight during the optimization process.

**4.2 The Mg-Zn binary system**

In the Mg-Zn binary system, the parameter of impurity diffusion of Zn in hcp-Mg $\Psi_{Zn}^{Mg}$, was directly obtained from the experimental data [14][17]. Among them, the values re-calculated in this work from the raw concentration profiles played a significant role in the optimization process. Due to the lack of experimental information about impurity diffusion of Mg in hcp-Zn, the end-member $\Psi_{Mg}^{Zn}$ was assumed to be equivalent to the self-diffusion coefficient of Zn. The binary interaction parameters

$^{0}\Psi_{Zn}^{Zn,Mg}$ and $^{1}\Psi_{Zn}^{Zn,Mg}$ were assessed according to the experimental interdiffusion coefficients and the concentration profiles.

The predictions from the optimized mobility parameters of the evolution of the impurity diffusion of Zn in hcp-Mg with the temperature is compared with the experimental data from the literature [14][17] in Fig. 3. They can be expressed as an Arrhenius-type equation, which depends on the frequency factor $D_0$ and the activation energy $Q$, which were given by 8.68*10$^{-4}$ m$^2$/s and 138.74 kJ/mol, respectively. The results of the optimization are in good agreement with the critically reviewed experimental values. The calculated values of the interdiffusion coefficients of the hcp Mg-Zn in the temperature range 623K-723K are plotted in Fig. 4 as a function of the Zn content together with the experimental values from [15]. They are generally in good agreement although the experimental values show a little valley at 673K, which means that the diffusion coefficients decreases as the Zn content increases. This is not reasonable, as indicated above, thus the weight of this data was reduced in the optimization process.

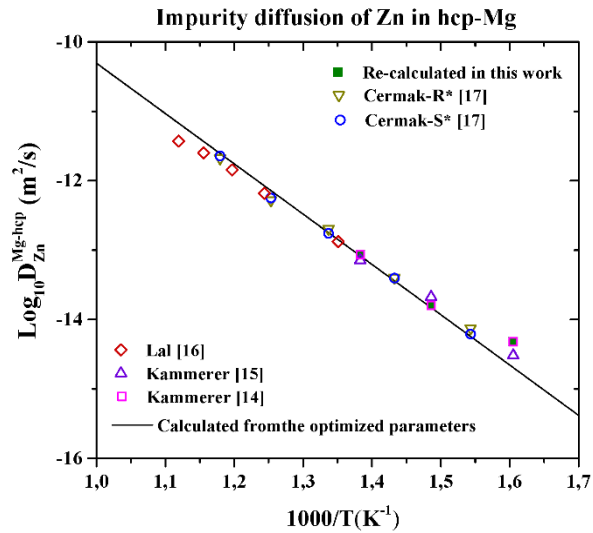

Figure 3. Variation of the Zn impurity diffusion coefficient in pure Mg ($D_{Zn}^{Mg}$) as a function of the inverse of the absolute temperature. The solid line stands for the predictions with the optimized parameters, which is compared with the experimental data in the literature [14][17] as well as with the data re-calculated in this work from the concentration profiles.

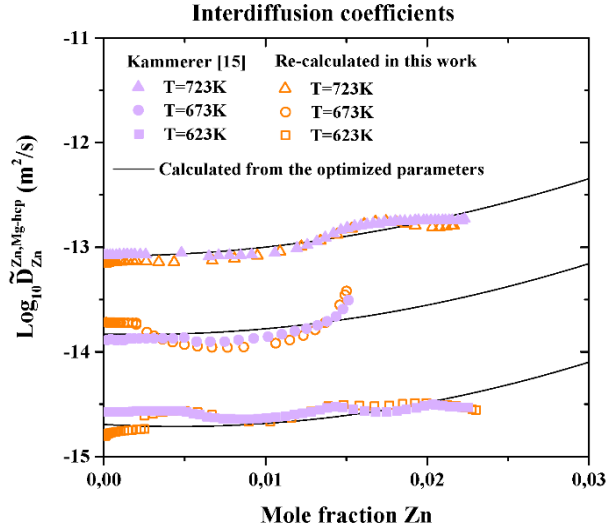

Figure 4. Interdiffusion coefficients of the hcp Mg-Zn couple ($Log D_{Zn}^{Zn,Mg}$) as a function of the Zn content at 623K, 673K and 723K. The solid lines stand for the predictions with the optimized parameters, which are compared with the experimental values from [15] as well as with the data re-calculated in this work from the concentration profiles.

### 4.3 The Zn-Al binary system

Owing to the anisotropy of Zn induced by the hcp structure, the self-diffusion coefficients, parallel to c-axis and perpendicular to c-axis, are different. They have been studied independently in the literature [23]-[25], and there is not an average self-diffusion coefficient. Thus, the end-member $\Psi_{Zn}^{Zn}$ was determined by Cui [26] based on previous experimental data, which was directly used in this work. In addition, there are not reliable experimental results for self-diffusion of Al in hcp-Al and for impurity diffusion of Al in hcp-Zn, because there is no stable Al-hcp structure and it is hard to melt high purity Zn due to its high vapor pressure. Thus, semiempirical self-diffusion relations have been used to estimate the end-member $\Psi_{Al}^{Al}$, following the theoretical estimations by Cui [26]. The end-member $\Psi_{Al}^{Zn}$, was determined based on the experimental data of the impurity diffusion coefficients of Ga in hcp-Zn [27], assuming that Ga has the similar properties with Al from the diffusion viewpoint.

Moreover, the parameters parallel to the c-axis were selected in the present work as the average values in the cases where average values were not available. Since the current results were focused in the region with high Mg content, the parameters of the region with high Zn content have limited influence. There is no available experimental study for hcp phase in Zn(Al) solid solution due to the small solubility of Al in hcp-Zn. Thus, the interaction parameters for this binary system are missing.

**4.4 The Mg-Al-Zn ternary system**

The parameters $\Psi_{Zn}^{Al,Mg}$ and $\Psi_{Al}^{Zn,Mg}$ were assessed from the raw concentration profiles and the impurity diffusion coefficients of Zn in Mg-Al alloys and Al in Mg-Zn alloys extracted in this work from the concentration profiles using the Hall method. The ternary interaction parameters play a negligible role in atomic mobility on account of the limited solubility of Al and Zn in hcp Mg alloy, and were assumed to be zero.

The predictions based on the optimized mobility parameters of the impurity diffusion coefficients of Al in Mg-Zn alloy and of Zn in Mg-Al are plotted as a solid line in Figs. 5a and 5b, respectively, together with the calculated values obtained from the original concentration profiles as well as from the available experimental data in literature [14][17]. It is evident that there is a large deviation in the predictions of the Al impurity diffusion coefficients, especially at lower Zn composition, as shown in Fig. 5(a), which may be due to errors introduced during the calculation by the Hall method. The calculated Zn impurity diffusion coefficients in Mg-Al alloy are plotted as solid lines in Fig. 5b and compared with the experimental data obtained from the diffusion couples [14][17]. The agreement between both is very good although the experimental values [17] lead to slightly higher coefficients than the calculated results.

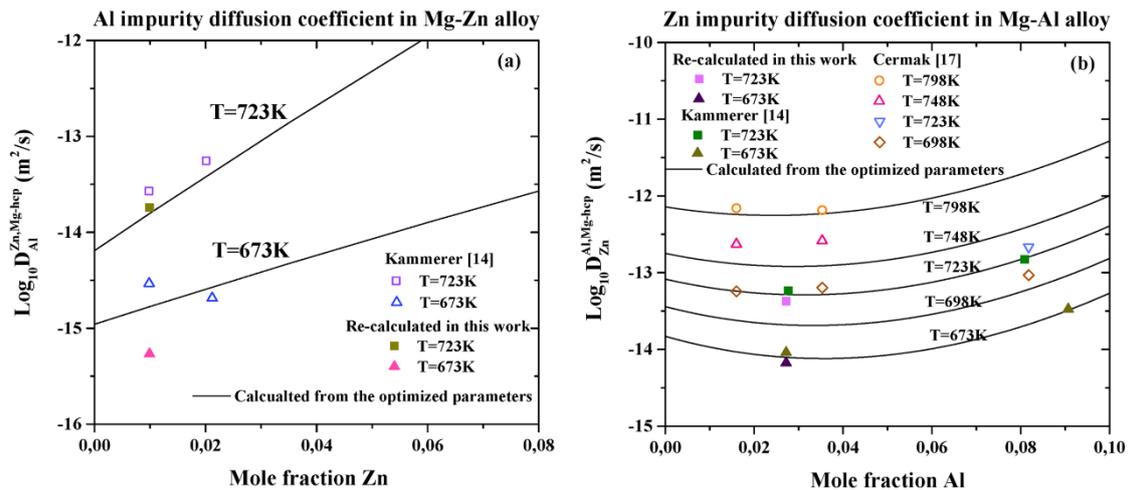

Figure 5. The experimental impurity diffusion coefficients of (a) Al in hcp Mg-Zn alloy ($D_{Al}^{Zn,Mg}$) with the experimental values [14]; (b) Zn in Mg-Al hcp alloy ($D_{Zn}^{Al,Mg}$) with the experimental values [14][17]

compared with the re-calculated values from the concentration profiles shown in solid symbols. The solid lines stand for the predictions with the optimized parameters.

The interdiffusion coefficients at 673K and 723K at common intersection compositions were obtained using the Whittle-Green method based on the experimental concentration profiles, and they are presented in Table 3 (Fig. 6). Zn diffuses much faster than Al, since the main-diffusion coefficient of Zn is one order of magnitude greater than that of Al. There appears a relative maxima in the profiles, where the gradient of Al goes up to zero, while that of Zn does not. Based on "Darken-type" couple method, one main-interdiffusion coefficient can be determined at the relative maxima, and it is reported in Table 4 (Fig. 6).

Table 3. Interdiffusion coefficients ($D \times 10^{-15} m^2/s$) at 673K and 723K extracted from the diffusion couple experiments of Kammerer [14]

| Temp | Couple | Al.at% | Zn.at% | $\widetilde{D}_{Al,Al}^{Mg}$ | $\widetilde{D}_{Al,Zn}^{Mg}$ | $\widetilde{D}_{Zn,Zn}^{Mg}$ | $\widetilde{D}_{Zn,Al}^{Mg}$ |
|---|---|---|---|---|---|---|---|
| 673K | II-III | 2.68 | 0.28 | 0.63 | 0.11 | 10.90 | 14.40 |
|  | II-IV | 0.80 | 0.63 | 0.57 | 1.60 | 37.95 | 3.54 |
| 723K | II-III | 2.75 | 0.24 | 25.71 | 3.85 | 70.19 | 15.85 |

Table 4. Main-interdiffusion coefficients of Zn ($D \times 10^{-14} m^2/s$) at relative maxima compared with the values obtained from the optimized mobility parameters

| Temp | Couple | Al.at% | Zn.at% | $\widetilde{D}_{Zn,Zn}^{Mg}$ experiment | $\widetilde{D}_{Zn,Zn}^{Mg}$ optimized |
|---|---|---|---|---|---|
| 673K | I | 9.06 | 0.10 | 3.52 | 3.58 |
|  |  | 0.00 | 2.15 | 1.97 | 3.15 |
|  | II | 0.02 | 0.90 | 1.79 | 1.62 |
|  |  | 2.70 | 0.07 | 0.60 | 0.81 |
|  | III | 2.68 | 0.31 | 0.95 | 0.88 |
|  |  | 0.00 | 0.07 | 1.38 | 1.48 |
|  | IV | 0.87 | 0.89 | 3.22 | 1.34 |
|  |  | 0.00 | 0.25 | 5.24 | 1.48 |
| 723K | I | 9.10 | 0.12 | 13.48 | 28.18 |
|  |  | 0.02 | 1.86 | 15.64 | 16.07 |
|  | II | 2.77 | 0.03 | 6.31 | 5.25 |
|  |  | 0.00 | 0.96 | 11.62 | 9.93 |
|  | III | 2.79 | 0.30 | 5.26 | 5.84 |

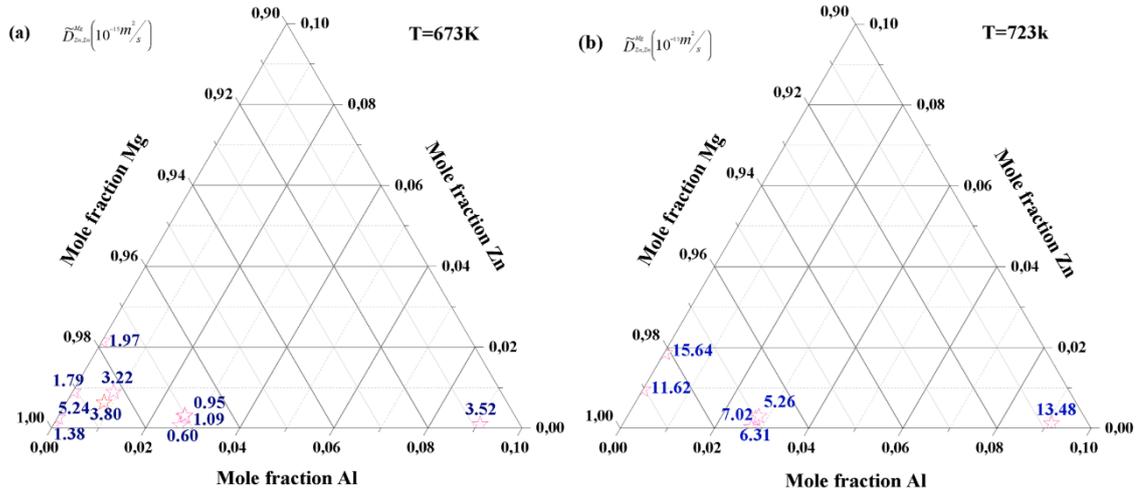

Figure 6. The main-interdiffusion coefficients of Zn ($D_{Zn,Zn}^{Mg}$) in the hcp Mg-Al-Zn alloys (a) at 673k; (b) at 723K

## 5. Diffusion Simulation

The set of optimized mobility parameters in Table 2, together with corresponding thermodynamic database, is able to predict the concentration profiles and the diffusions paths, as shown below. This exercise further validates the assessed parameters.

### 5.1 Concentration profiles of diffusion couples

The concentration profiles for the Mg-Al binary system calculated with the optimized mobility parameters are compared in Fig. 7 with the raw experimental data obtained from the diffusion couples from the literature [15]. The agreement is very good at 673K and 723K, but there are large differences at 623K, which may be due to the discrepancy in the interdiffusion coefficients curves, as indicated in §4.1. Similarly, the calculated and experimental concentration profiles in Mg-Zn alloy are plotted in Fig. 8 and they are also in good agreement. The current simulated results are in good accordance with the corresponding experimental data [15]. As it can be seen at 673K, there is a sharp decrease in the concentration profile on the side of Mg(Zn) solid solution. This is not reasonable, as it was already indicated in section §4.2.

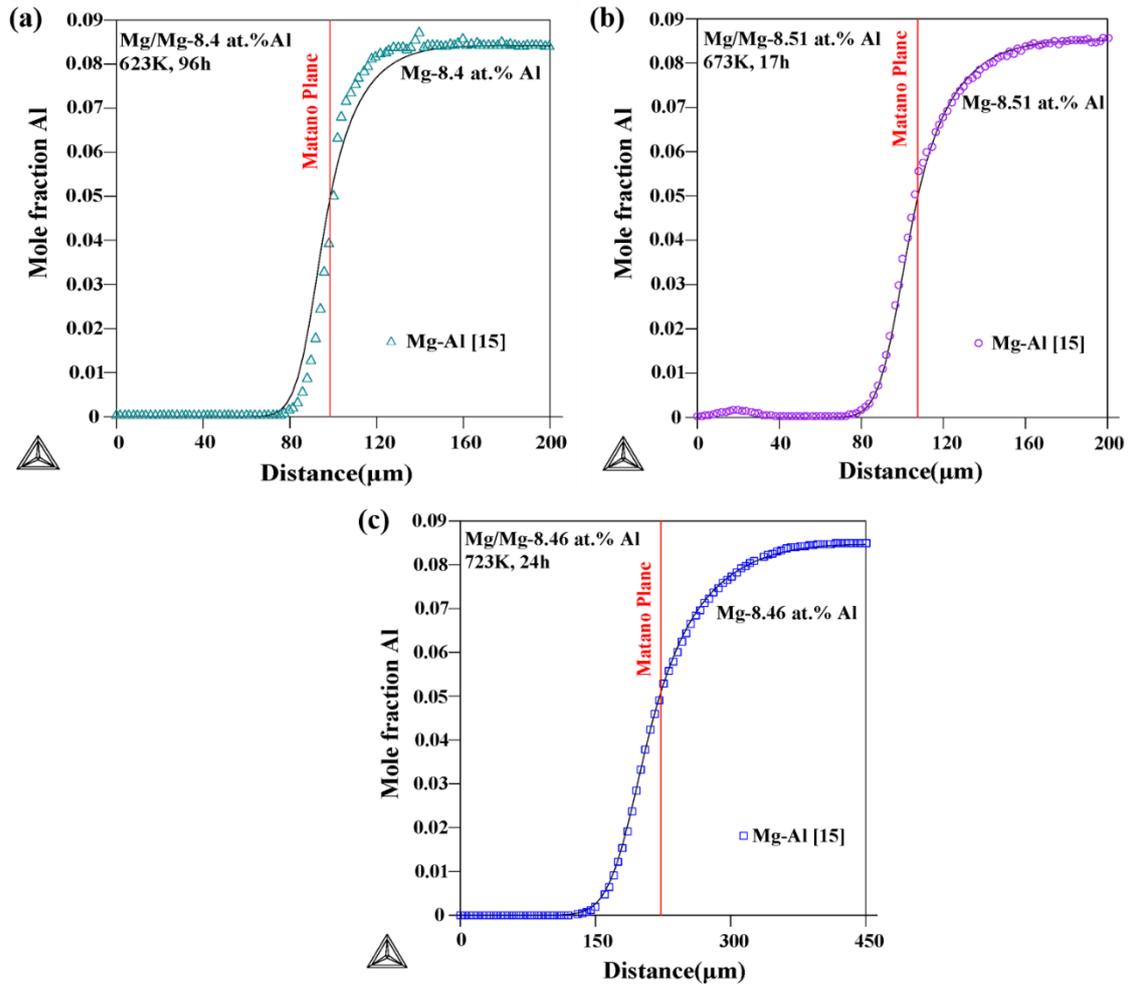

Figure 7. Concentration profiles calculated from the optimized parameters and experimental values [15] of diffusion couples in Mg-Al: (a) at 623K; (b) at 673K; (c) at 723K. The solid lines stand for the predictions with the optimized parameters.

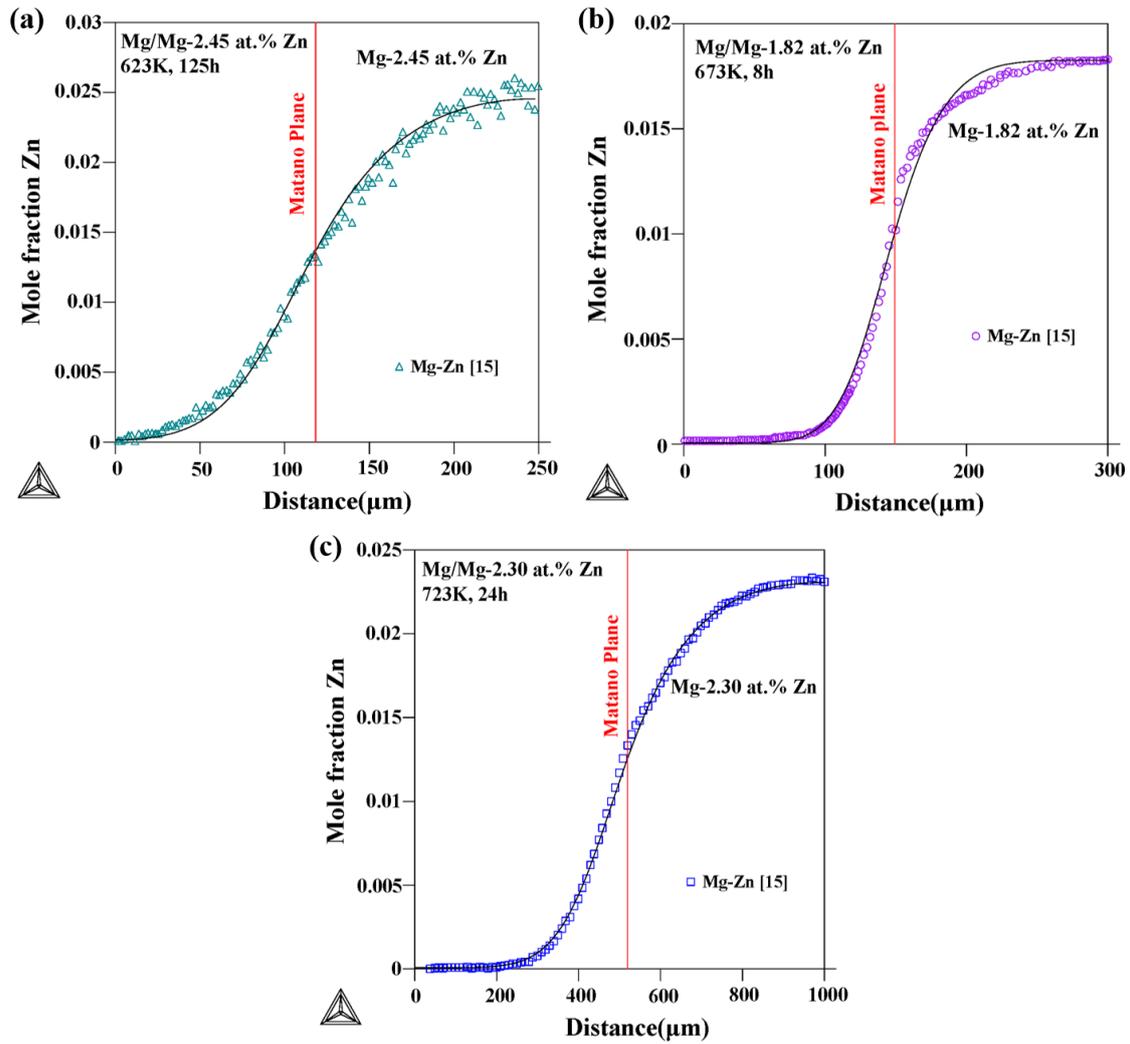

Figure 8. Concentration profiles calculated from the optimized parameters and experimental values [15] of diffusion couples in Mg-Zn: (a) at 623K; (b) at 673K; (c) at 723K. The solid lines stand for the predictions with the optimized parameters.

Representative concentration profiles calculated from the optimized parameters for the ternary couples are plotted in Fig. 9, together with the experimental data. Figure 9(a) and (b) show the calculated concentration profiles of the couple I (Mg-9at.%Al/Mg-3at.%Zn annealed at 673K for 8 hours) and of the couple II (Mg-3at.%Al/Mg-1at.%Zn annealed at 723K for 5 hours), along with the raw EPMA experimental data [18]. Overall, the optimized concentration profiles fit well with the experimental values in these cases. The calculated and experimental concentration profiles for the diffusion couple IV (Mg/Mg-1at.%Al-1at.% Zn annealed at 673K for 24 hours) are plotted in Fig. 9c and there are large differences in the concentration profiles of Zn. As shown in Table 4, the calculated main-interdiffusion coefficients for Zn of couple IV are about four times higher than those obtained from the experimental data of couple III. These differences in the interdiffusion coefficients with similar compositions are not reasonable and the experimental data for couple IV were not used in the optimization.

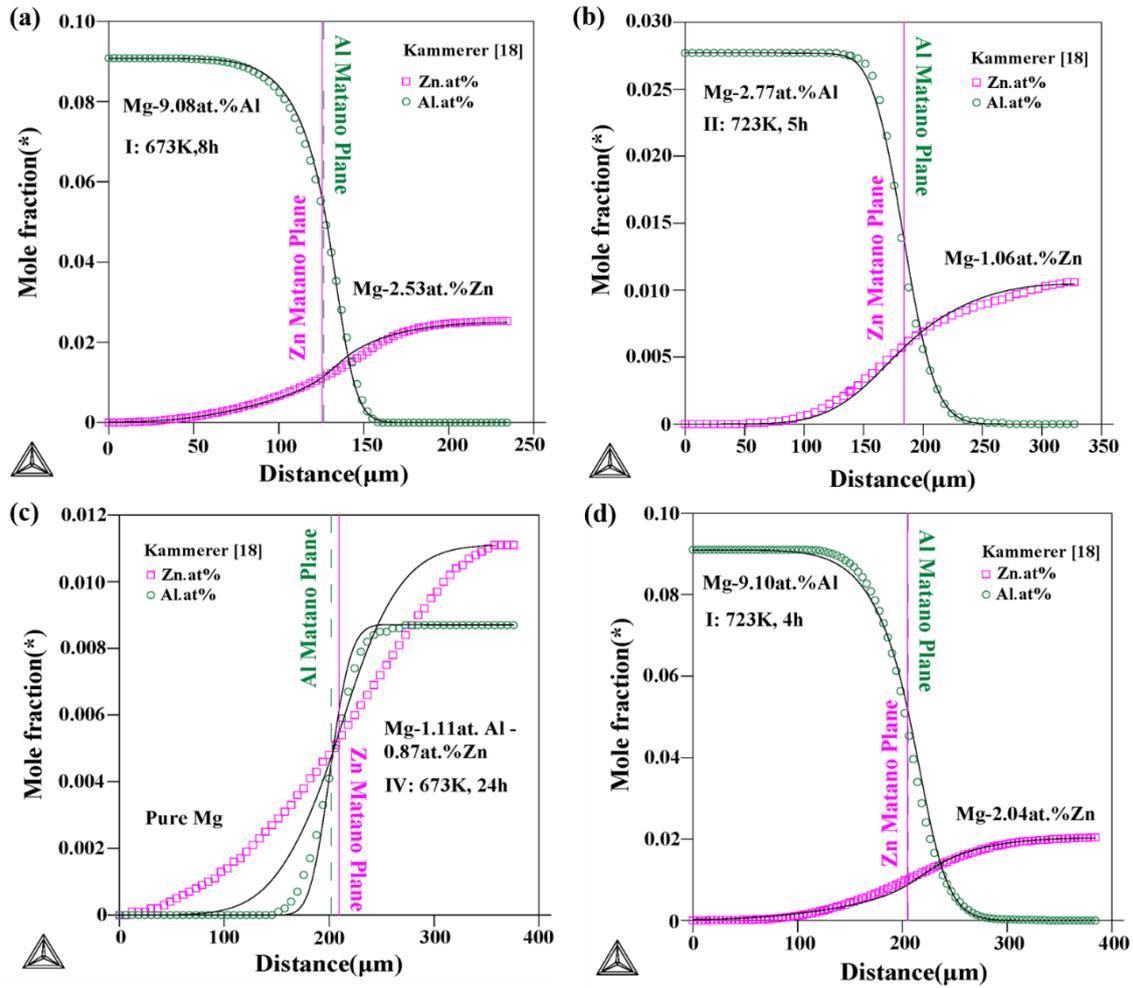

Figure 9. Concentration profiles calculated from the optimized parameters and experimental values [18] in hcp Mg-Al-Zn alloys. (a) Couple I annealed at 673K for 8 hours, (b) Couple II annealed at 723K for 5 hours, (c) Couple IV annealed at 673K for 24 hours, (d) Couple I annealed at 723K for 4 hours. The solid lines stand for the predictions with the optimized parameters.

The simulated concentration profile of diffusion couple I annealed at 723K is plotted in Fig. 9(d), and discrepancies were found at the terminal end of Al concentration profile. The deviation results from the impurity diffusion coefficient of Al in Mg-Zn alloy that should be adjusted consistently with other experimental data. Nevertheless, these differences are acceptable taking into account the general agreement with the experimental data. It can be clearly seen from the Fig. 9 that Zn diffuses faster than Al in the ternary system according to the longer diffusion distance of Zn in the solid solutions. The good agreement between the simulated concentration profiles and the experimental results validate the assessed atomic mobility in the current work.

## 5.2 Diffusion paths of the ternary diffusion couples

Figure 10(a) and (b) depict the diffusion paths calculated for the optimized parameters for various ternary diffusion couples annealed at 673K and 723K respectively, which are compared with corresponding experimental data [18]. Note that the diffusion paths are expressed to be S-shaped curves. There is an empirical trend that the initial directions of the diffusion paths tend to be aligned with the line of constant composition of the slower diffusion element in the diffusion zone. Meanwhile, diffusion paths can distinguish the diffusion rates for different elements, based on the difference in the greater degree of curvature and the shape of the paths. The majority of the calculated diffusion paths show good agreement with the corresponding experimental values, with the exception of couple I at both temperatures. The departure of several points from the calculated curves comes from the slight deviation in the concentration profiles of couple I at the terminal end of the Zn profile at 673K and that of the Al profile at 723K. Regardless of this discrepancy, the optimized parameters provided acceptable predictions taking into account their ability to reproduce the overall diffusional behavior of the Mg-Al-Zn system.

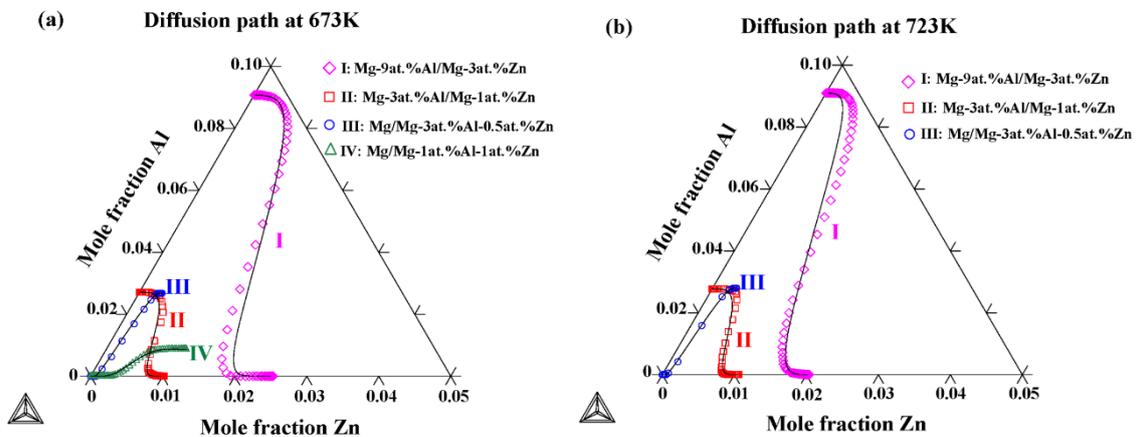

Figure 10. Comparison between the calculated diffusion paths and the available experimental data [18] (a) at 673K; (b) at 723K. The solid lines stand for the predictions with the optimized parameters.

## 6 Conclusions

The experimental data on diffusion of Mg-Al-Zn alloys available in the literature has been critically reviewed. Based on the concentration profiles from the literature, the diffusion coefficients have been re-extracted using the Hall method for impurity diffusion, and the Sauer-Freise and the Whittle-Green strategies for interdiffusion coefficients in binary and ternary systems, respectively. In addition, the concentration profiles with relative maxima were analyzed by the "Darken-type" couple method. On basis of these data, the atomic mobility of Al, Zn and Mg in hcp Mg-Al-Zn ternary alloys has been assessed, via the Parrot Model of DICTRA software package in conjunction with the corresponding thermodynamic database. The optimized set of atomic mobility parameters could successfully predict the diffusion behavior during binary-couple and ternary-couple experiments. Comprehensive comparisons between the simulated and the measured diffusional data showed an excellent agreement not only for the concentration profiles but also for the diffusion paths, hence further validating the assessed parameters.

## Acknowledgements

This investigation was supported by the European Research Council (ERC) under the European Union's Horizon 2020 research and innovation programme (Advanced Grant VIRMETAL, grant agreement No. 669141). Ms. Jingya Wang acknowledges the financial support from the China Scholarship Council. YC would like to thank the support of National Natural Science Funds of China (Grant No.51003034).

**Appendix: Determination of interdiffusion coefficients from "Darken-type" diffusion couple**

The diffusion couples that exhibit maxima and/or minima are termed "Darken-type" couples [28], and the extra interdiffusion coefficients can be determined from them. The phenomenological description of multicomponent diffusion originally proposed by Onsager [29][30], has been utilized for theoretical as well as experimental diffusion studies in multicomponent alloys. According to the Onsager's formalism [29][30], an extension of Fick's law has been demonstrated by Kirkaldy [31] in an n-component system given as following equation (1), with being referred to the laboratory-fixed frame with the molar volume of species $i$ being assumed to be constant. Meanwhile, the interdiffusion fluxes $J_i$ are determined directly from the concentration profiles of the individual components [32][33] as follow:

$$J_i = -\sum_{j=1}^{n-1} D_{i,j}^n \frac{\partial x_j}{\partial z} = \frac{1}{2t} \int_{x_i^- or x_i^+}^{x_i} (z - z_0) dx_i \quad (i = 1, 2 \ldots n) \tag{1}$$

where $x_i$ is the mole fraction of species $i$, then $x_i^+$ and $x_i^-$ stand for the mole fractions of $i$ at either end of the diffusion couple. Moreover, $z$ is the diffusion distance, and $z_0$ is the Matano plane for the diffusion couple. $t$ is the diffusion time, and $D_{i,j}^n$ is the interdiffusion coefficient with $n$ as the dependent species. The interdiffusion fluxes and concentration gradients are evaluated at the common composition for the individual components from their concentration profiles for the pairs. From (1), four independent equations are obtained to determine the two main-interdiffusion coefficients and the two cross interdiffusion coefficients at the intersection common composition.

In order to avoid the inaccuracy induced by the position of the Matano plane, the Whittle-Green [34] (W-G) method is used. It introduces a normalized concentration variable $Y_i = \frac{x_i - x_i^-}{x_i^+ - x_i^-}$, and the integral on the left hand side of equation (1) can be re-written as:

$$\int_{x_i^-}^{x_i} z \cdot dx_i = \left(x_i^+ - x_i^-\right) \left[ (1 - Y_i) \int_{-\infty}^{z} Y_i dz + Y_i \int_{z}^{+\infty} (1 - Y_i) dz \right] \tag{2}$$

By combining the equations (1) and (2), the Fick's second law of diffusion can be expressed as:

$$\gamma_1 = \left[(1-Y_1)\int_{-\infty}^{z} Y_1 \cdot dz + Y_1 \int_{z}^{+\infty}(1-Y_1) \cdot dz\right] = -2t\left(D_{11}^3 \cdot \frac{dY_1}{dz} + D_{12}^3 \frac{(x_2^+ - x_2^-)}{(x_1^+ - x_1^-)} \cdot \frac{dY_1}{dz}\right) \quad (3\text{-a})$$

and

$$\gamma_2 = \left[(1-Y_2)\int_{-\infty}^{z} Y_2 \cdot dz + Y_2 \int_{z}^{+\infty}(1-Y_2) \cdot dz\right] = -2t\left(D_{22}^3 \cdot \frac{dY_2}{dz} + D_{21}^3 \frac{(x_1^+ - x_1^-)}{(x_2^+ - x_2^-)} \cdot \frac{dY_1}{dz}\right) \quad (3\text{-b})$$

In the case of "Darken-type" diffusion couple, one of the main and one of the cross-interdiffusion coefficients at the extreme position can be obtained from the two concentration profiles for the alloy elements from the single diffusion couple. For instance, if there is a relative maximum or minimum for the element 1, the concentration gradient of the element is zero, namely $\frac{dx_1}{dz} = 0$. and the equation (3-b) will become be simplified to:

$$D_{22}^3 = \frac{dz}{dY_2} \cdot \frac{1}{2t}\left[(1-Y_2)\int_{-\infty}^{z} Y_2 \cdot dz + Y_2 \int_{z}^{+\infty}(1-Y_2) \cdot dz\right] \quad (4\text{-a})$$

$$D_{12}^3 = \frac{(x_1^+ - x_1^-)}{(x_2^+ - x_2^-)} \frac{dz}{dY_2} \cdot \frac{1}{2t}\left[(1-Y_1)\int_{-\infty}^{z} Y_1 \cdot dz + Y_1 \int_{z}^{+\infty}(1-Y_1) \cdot dz\right] \quad (4\text{-b})$$

From this equation, $D_{22}^3$ and $D_{12}^3$ can be calculated from the single couple at the composition of the extrema in the profile of component 1. Similarly, $D_{11}^3$ and $D_{21}^3$ can be determined at the section where $\frac{dx_2}{dz} = 0$ if the element 2 also exhibits a maximum in the concentration profiles.

If the concentration gradient of one component at the end of the couple approaches zero, meanwhile that of the other component is not equal to zero, ($\frac{dx_1}{dz} = 0$, while $\frac{dx_2}{dz} \neq 0$), in that case, the composition of component 1 will go up to the terminal composition (namely, $x_1 \approx x_1^+ or x_1^-$), and the $Y_i$ or $1-Y_i$ will also be zero. Hence, only one main interdiffusion coefficient $D_{22}^3$, can be obtained by the single diffusion couple using equation (4-a), while the main-interdiffusion coefficient $D_{11}^3$ can be determined by the concentration profile of element 1 at the end of the couple of element 2.

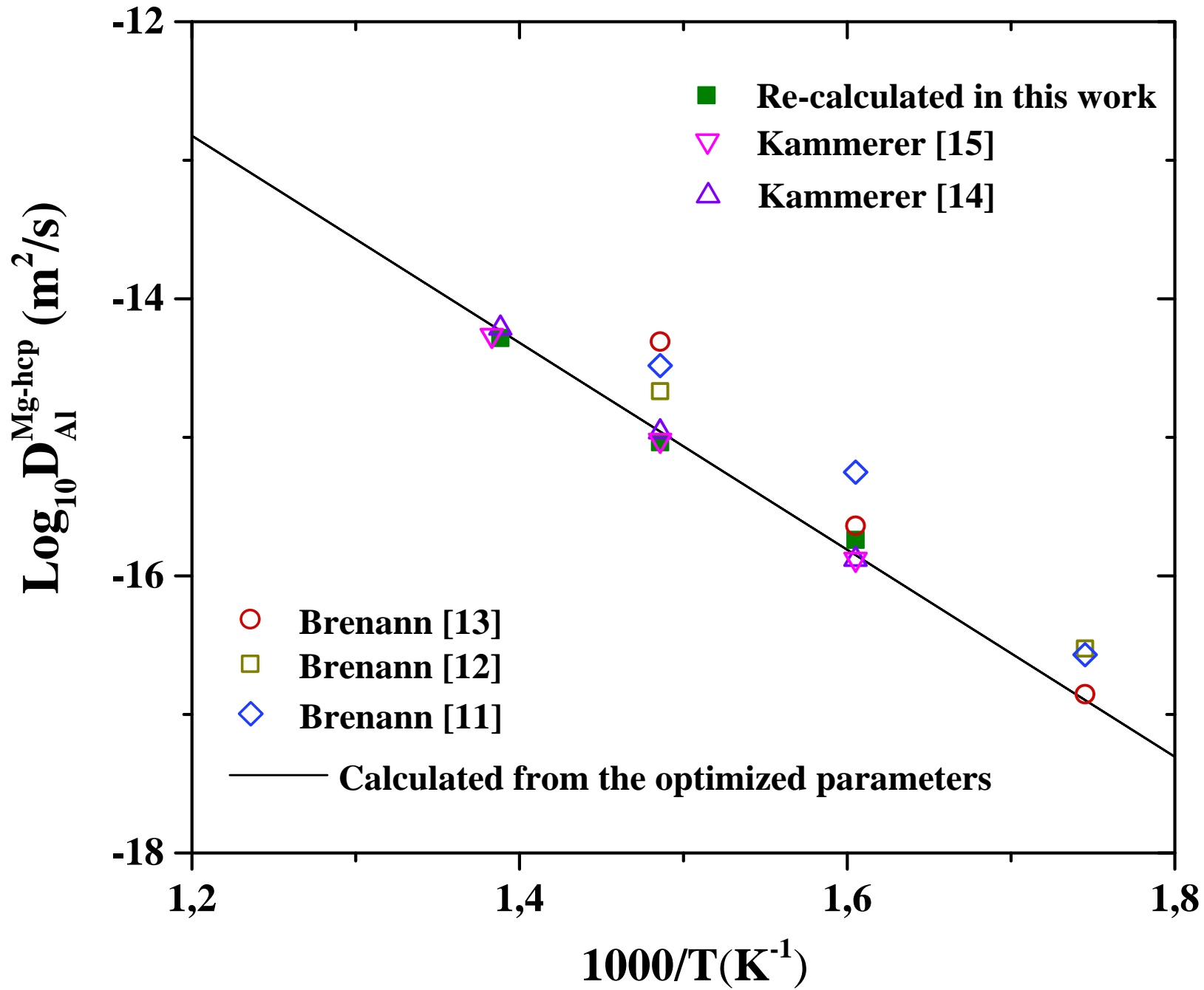

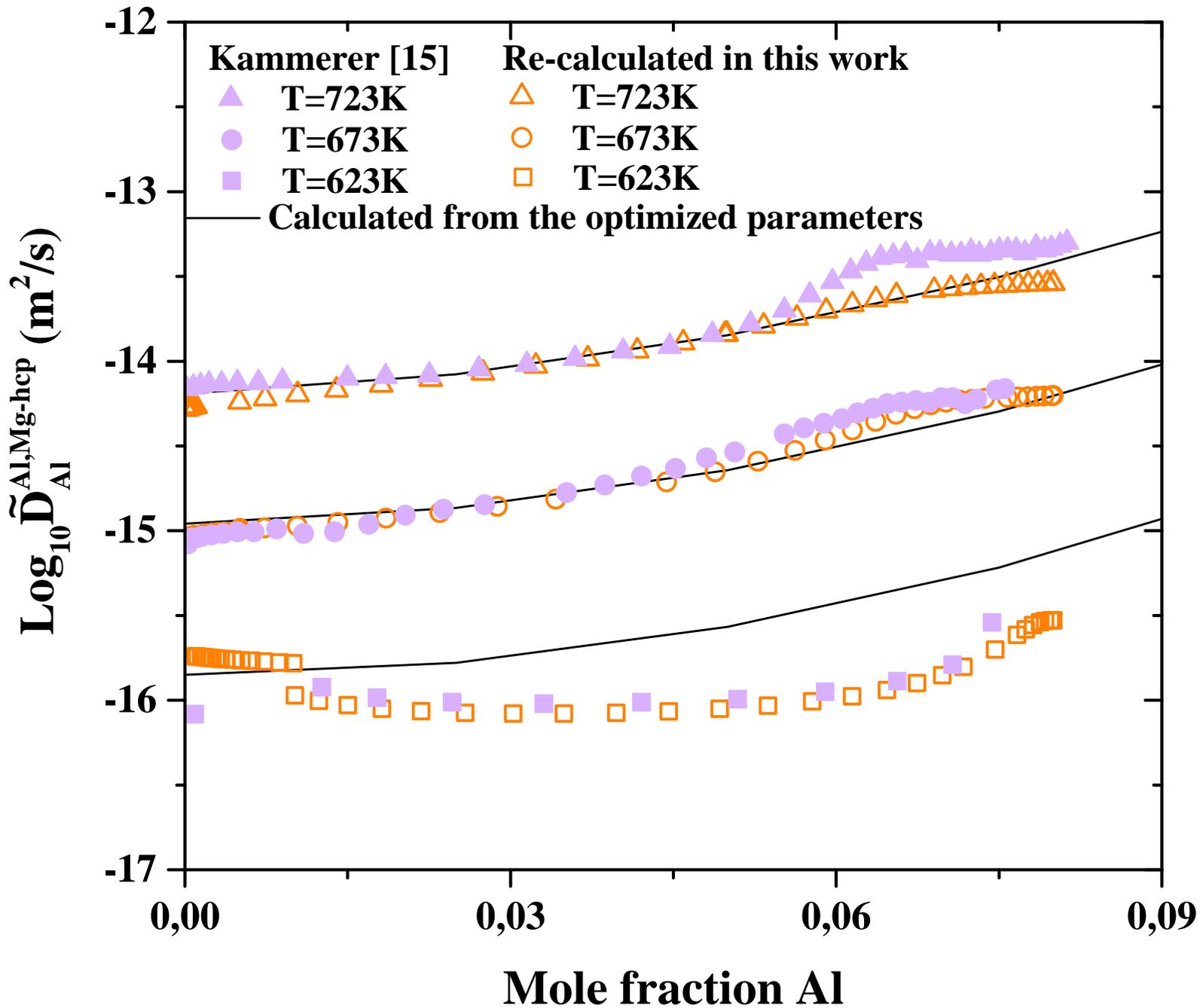

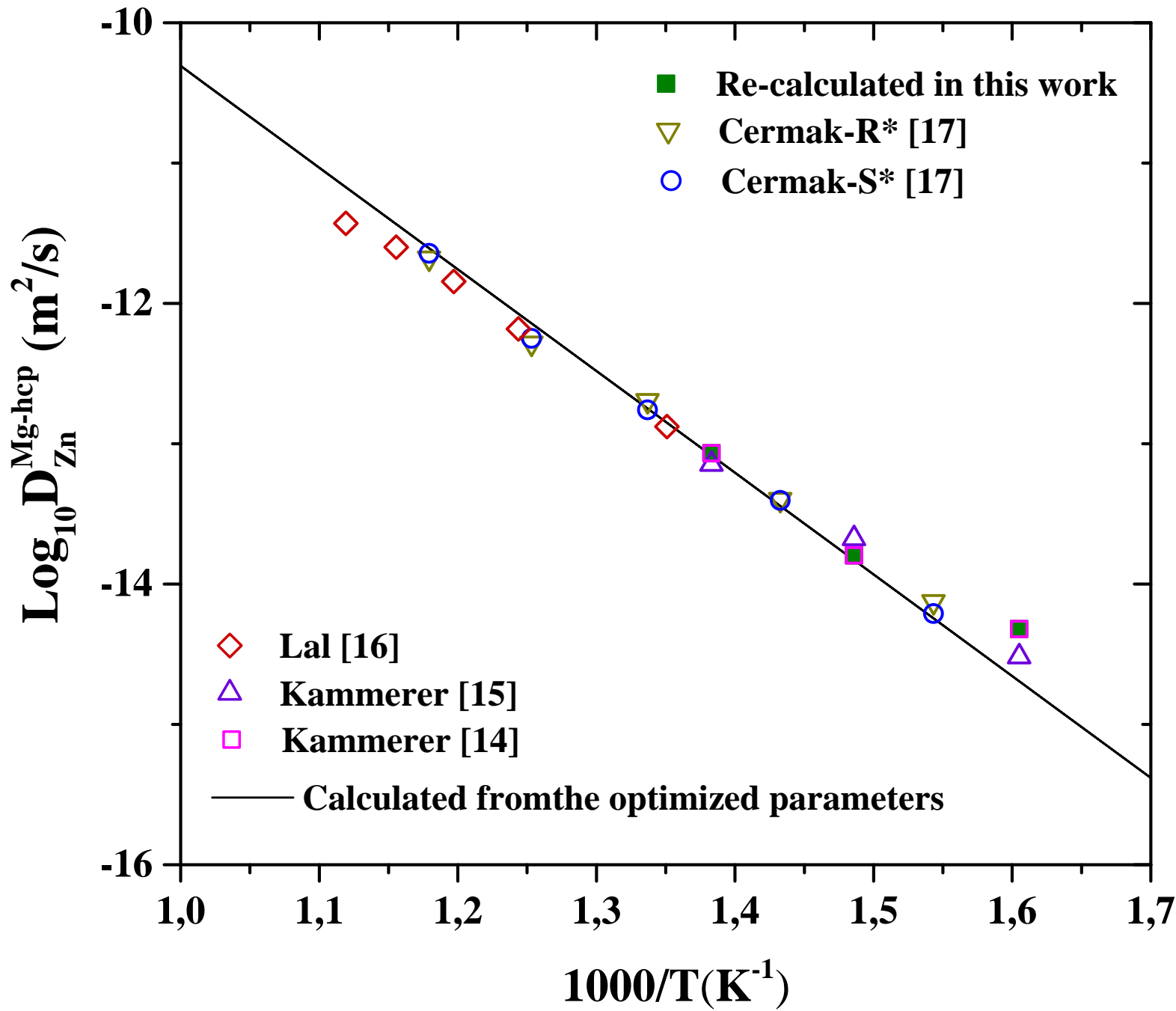

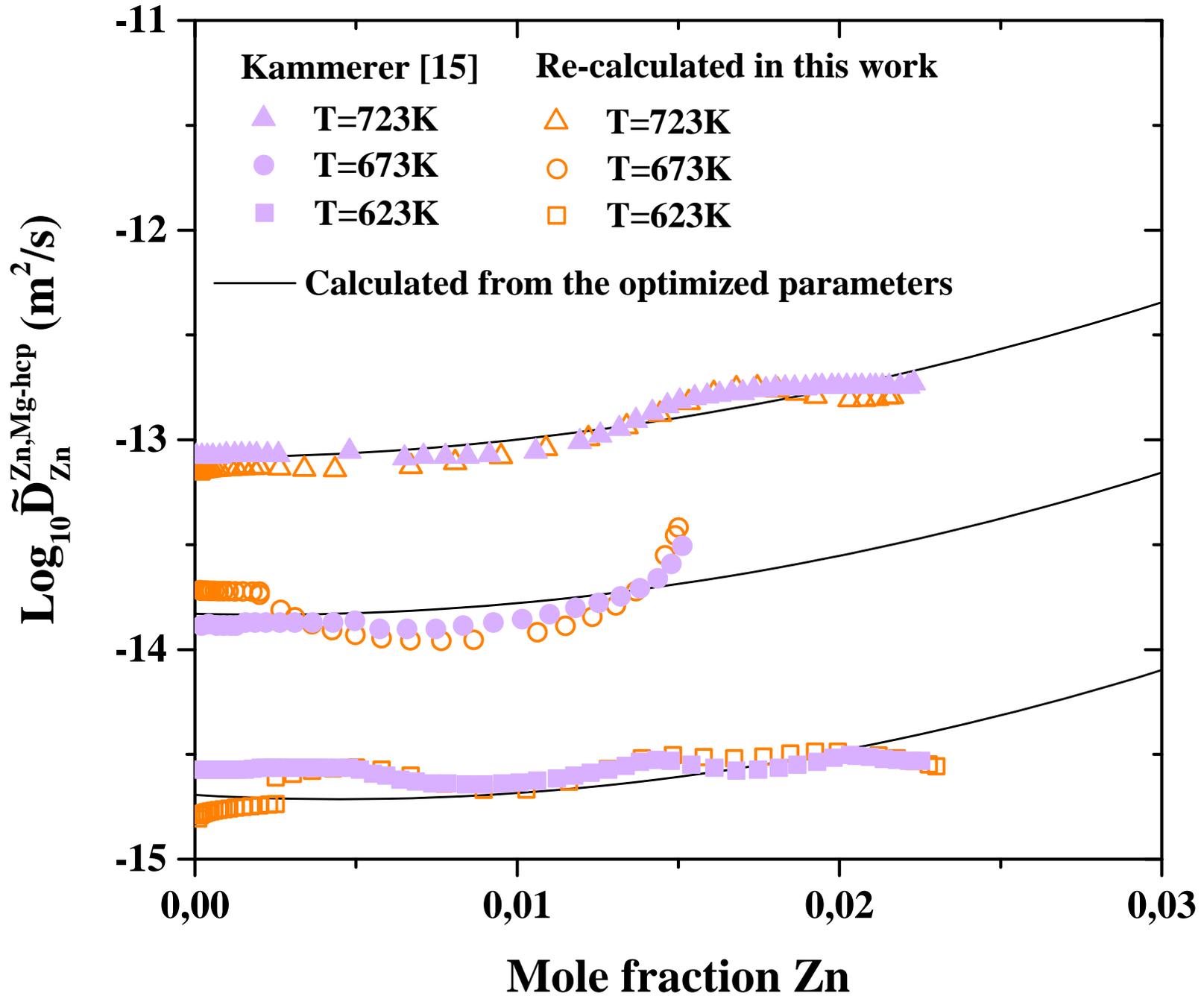

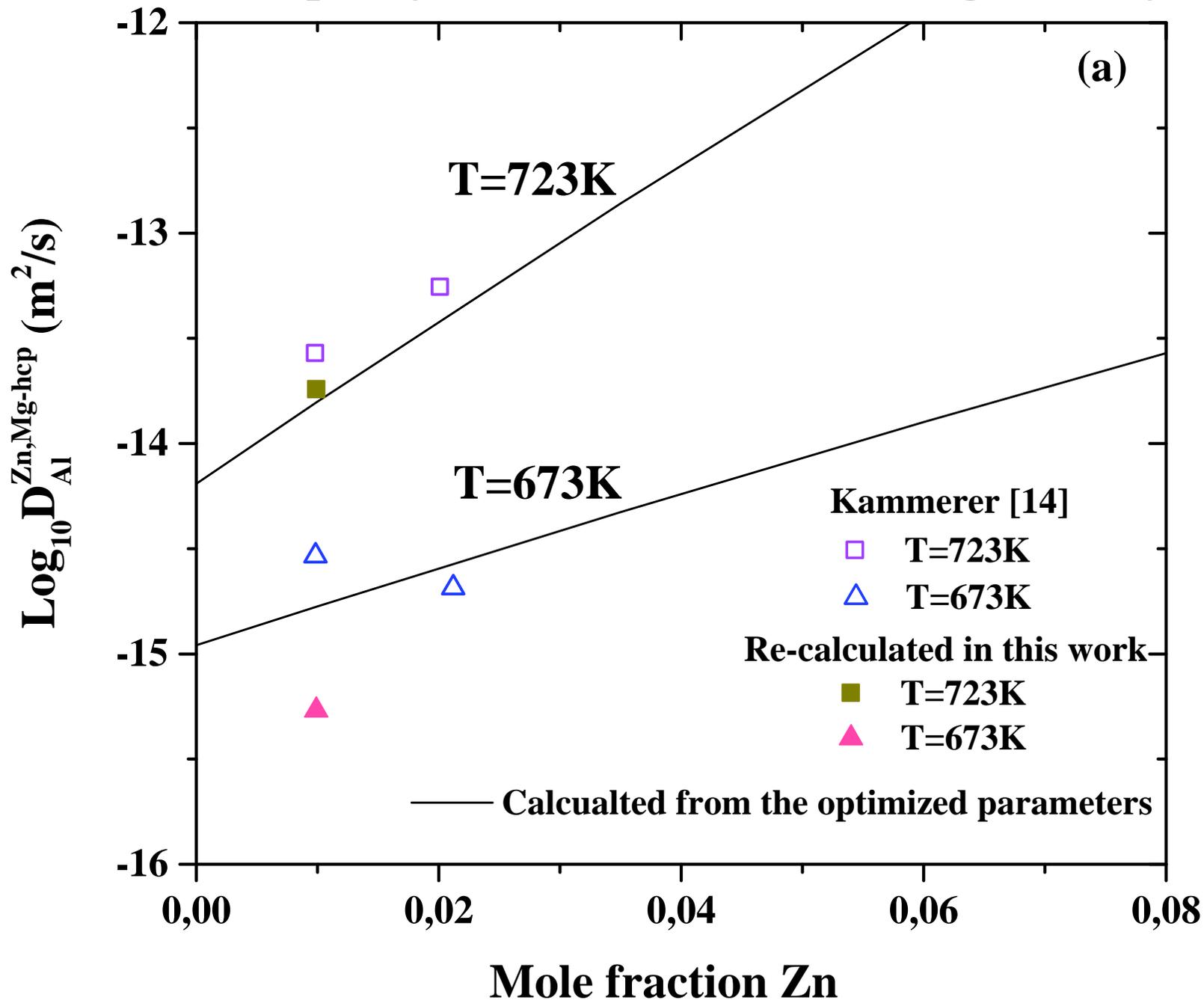

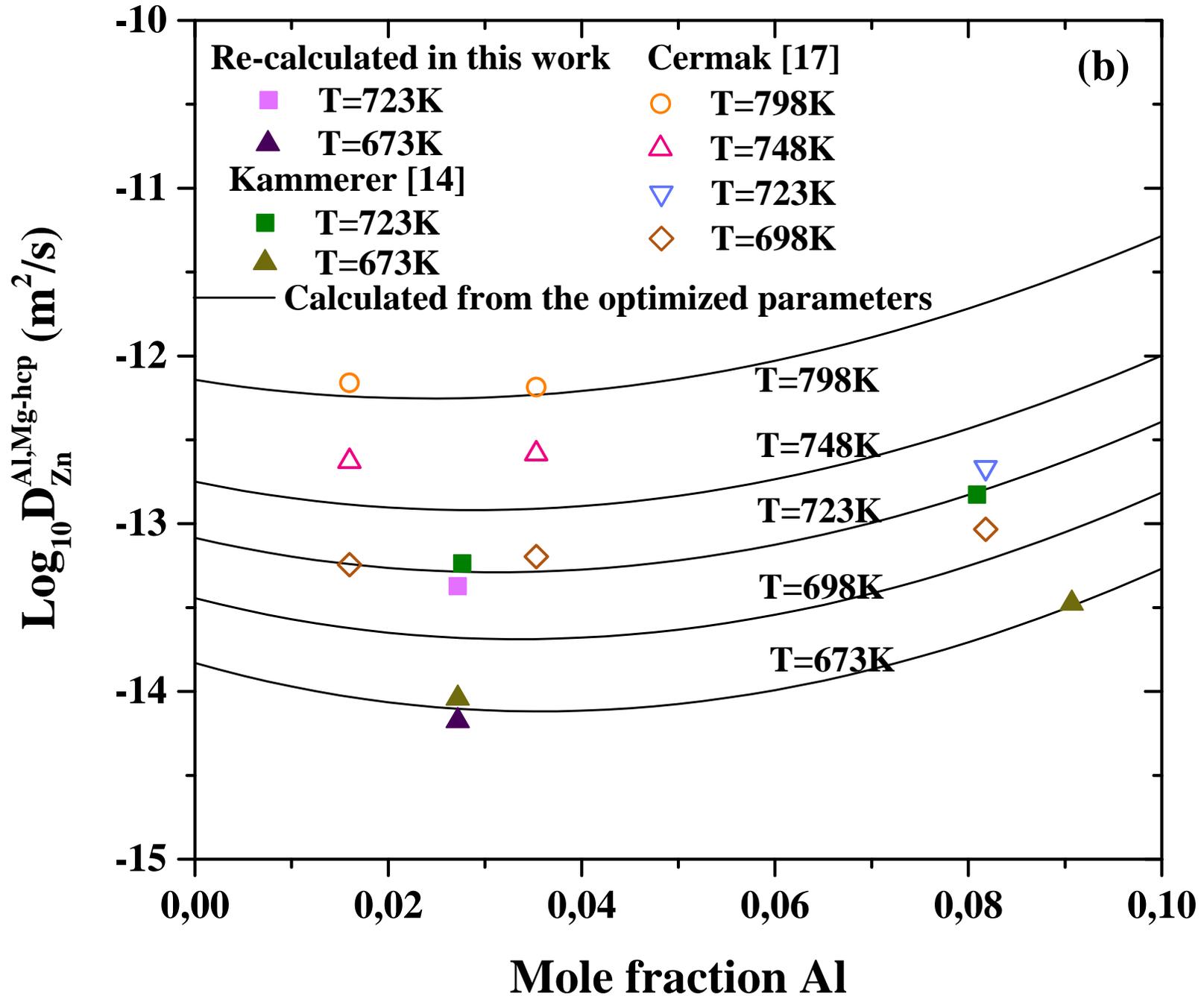

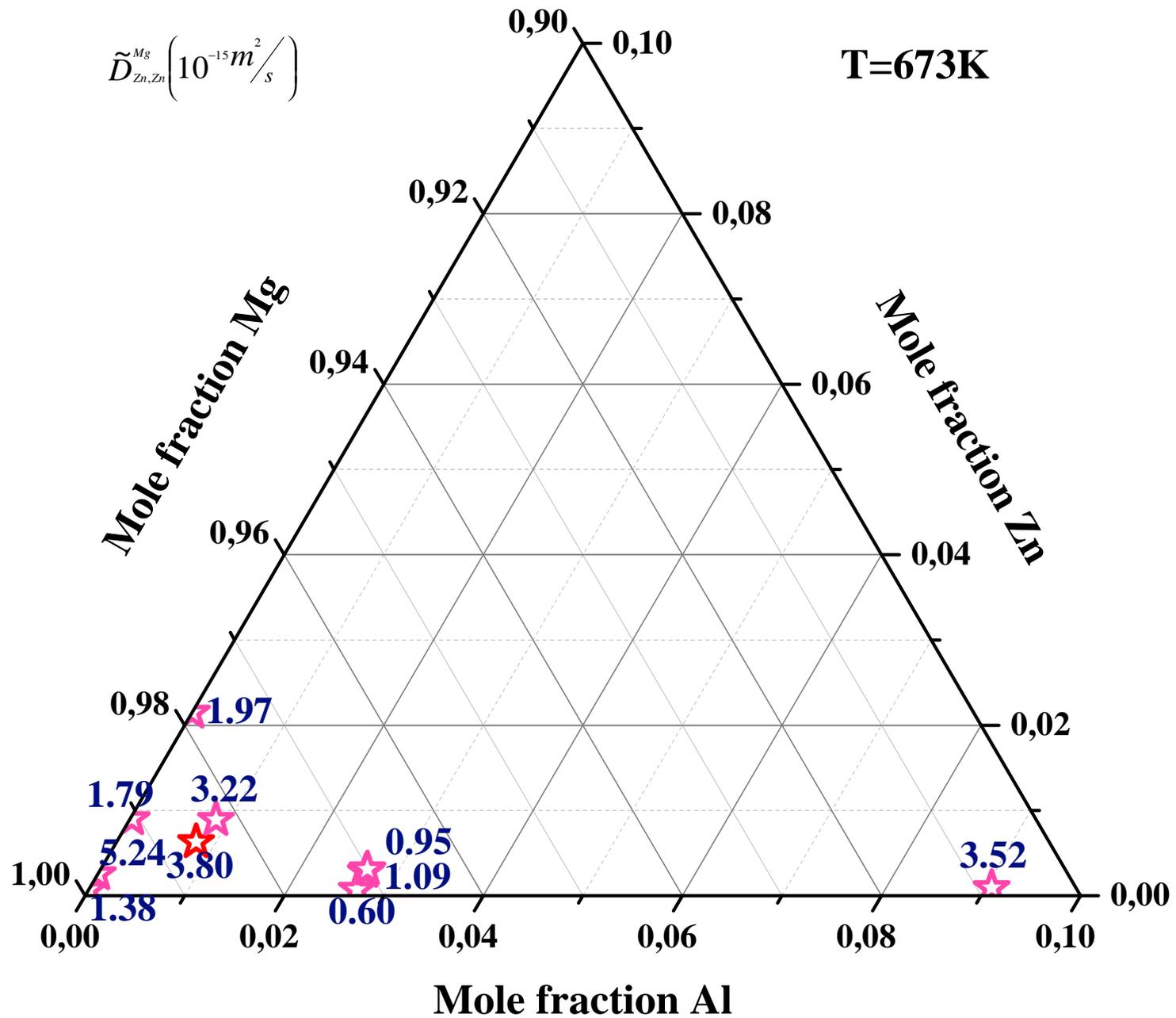

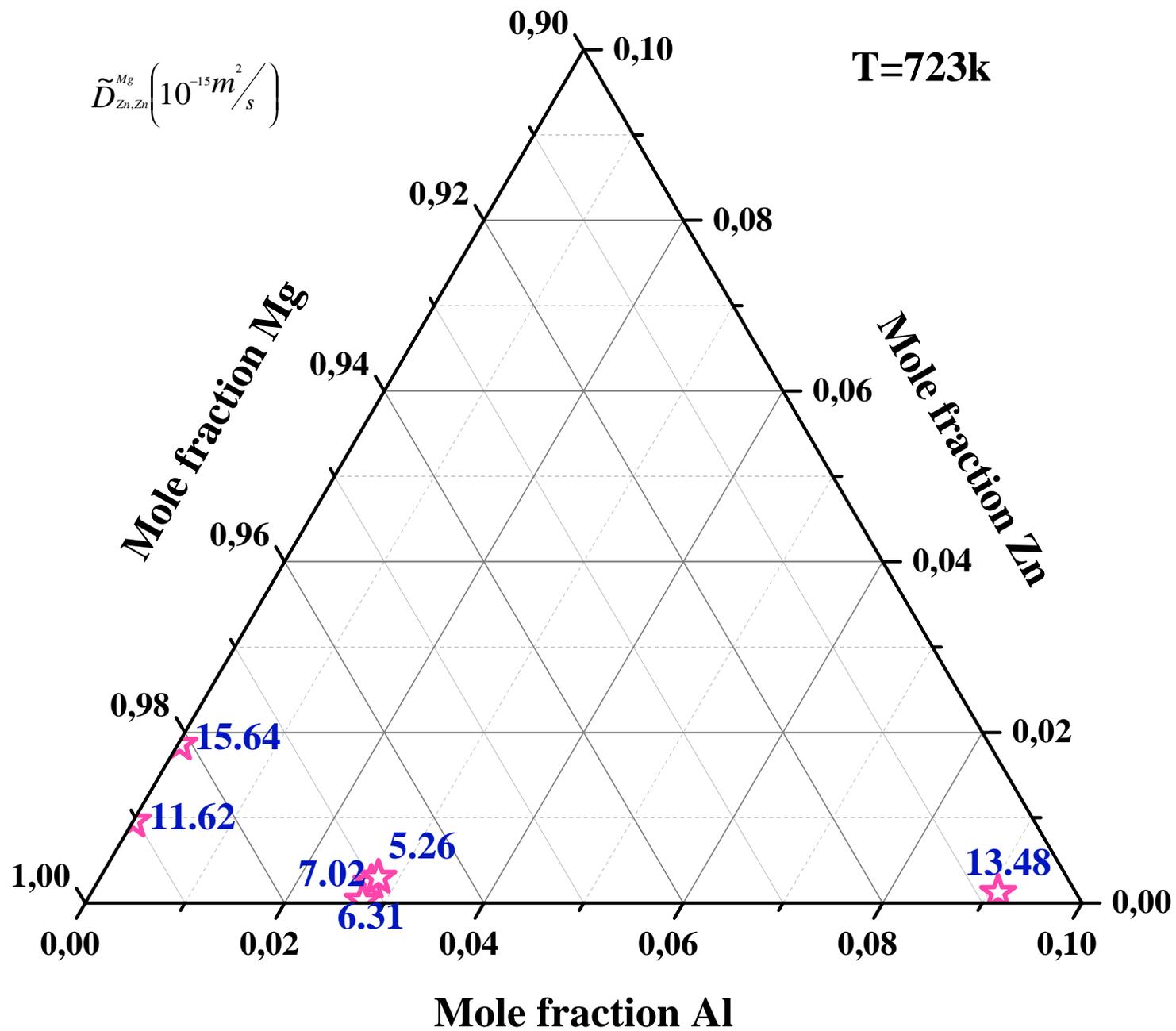

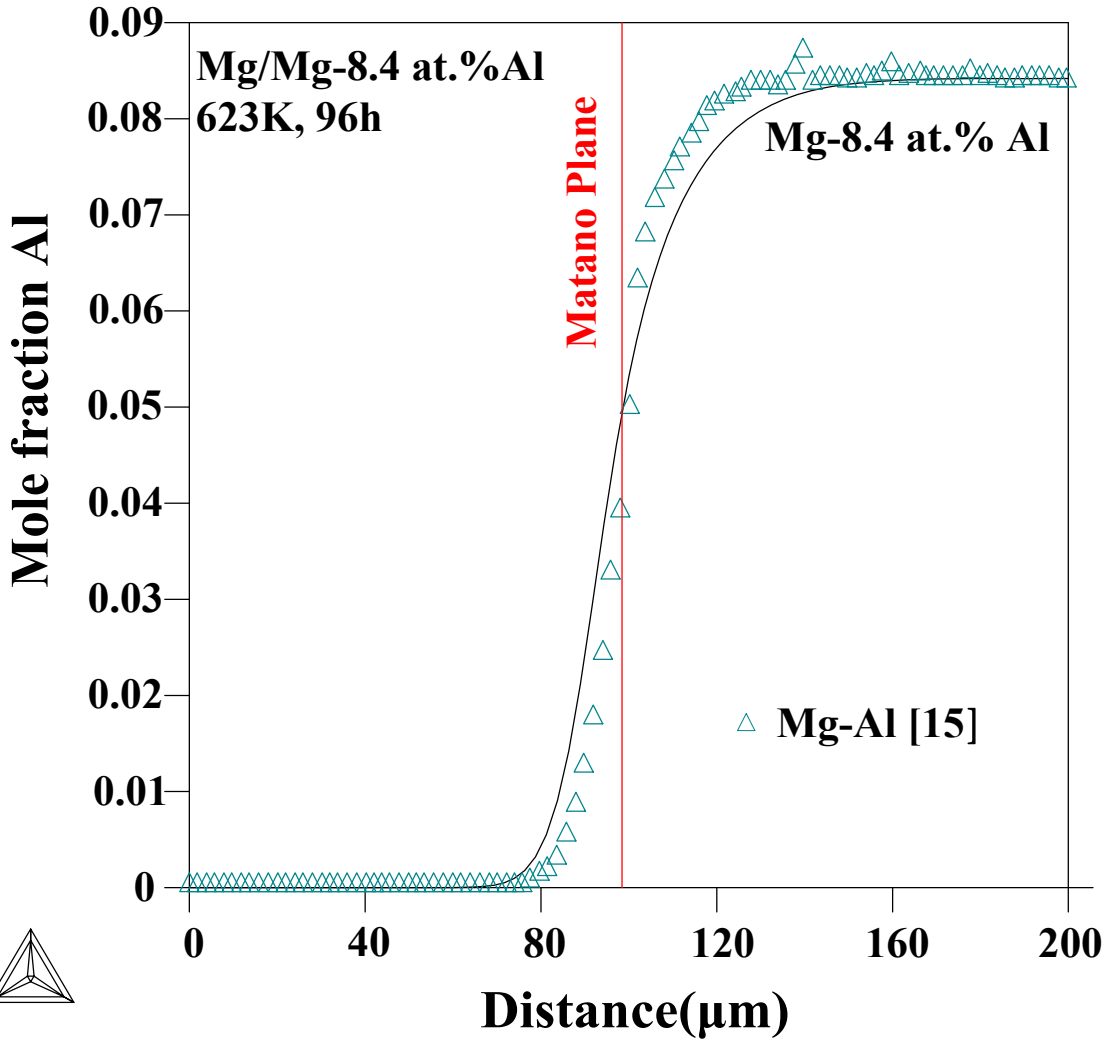

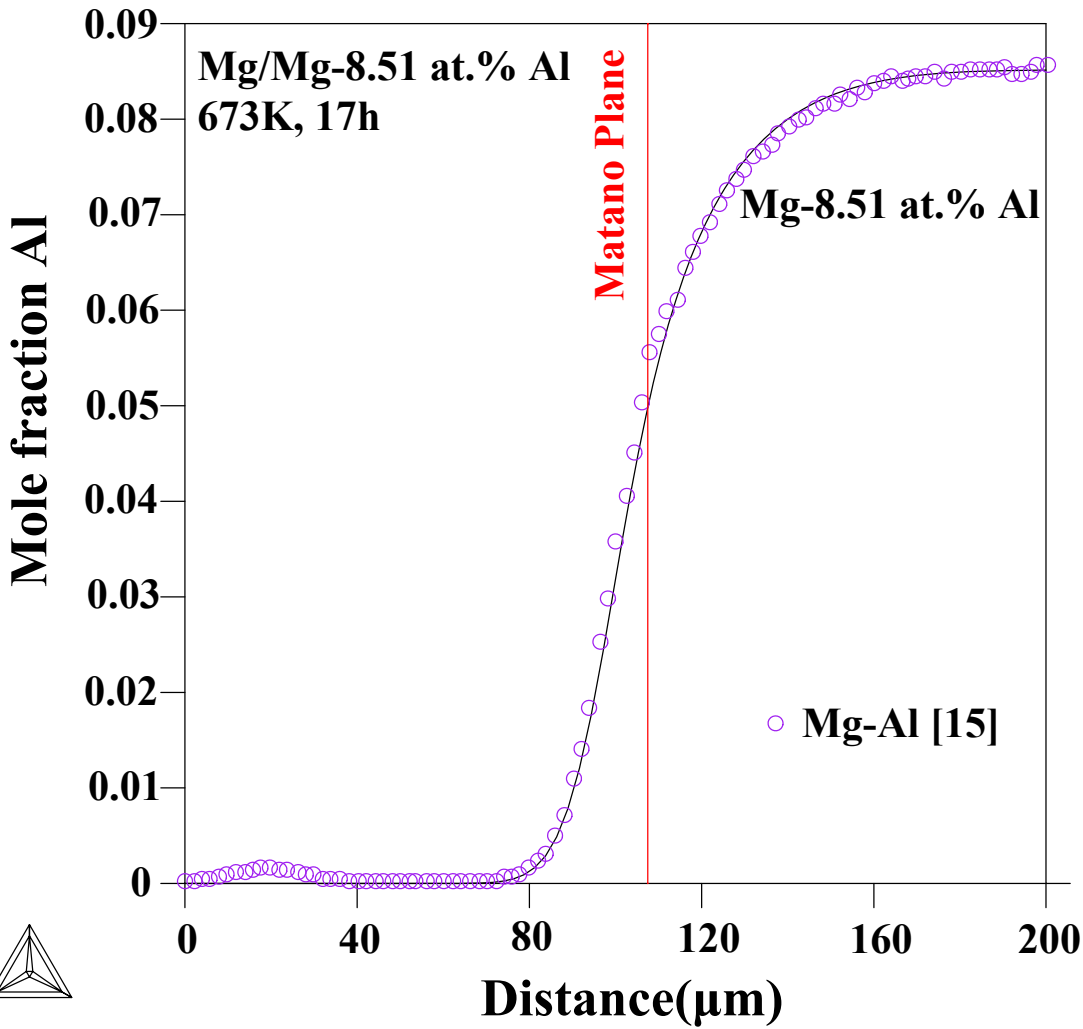

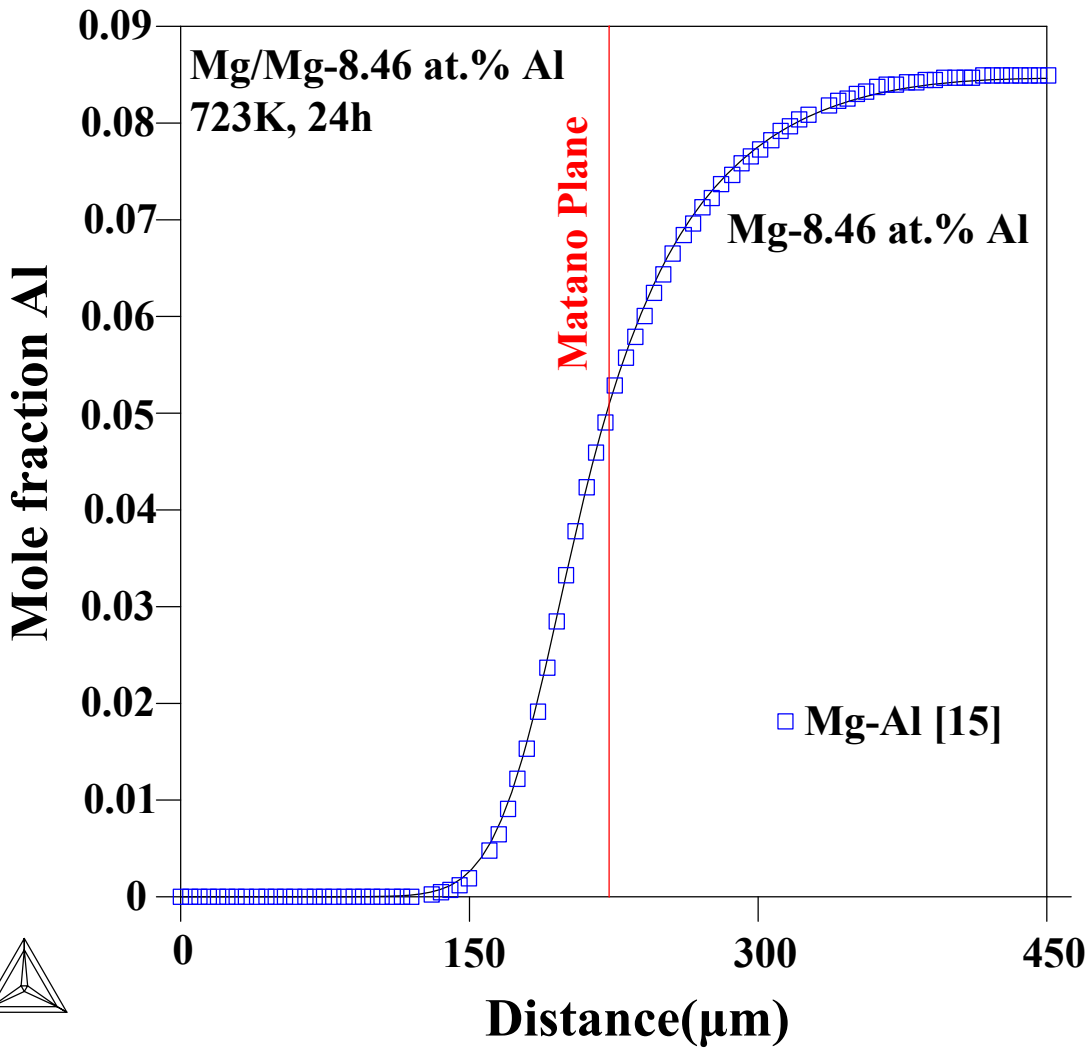

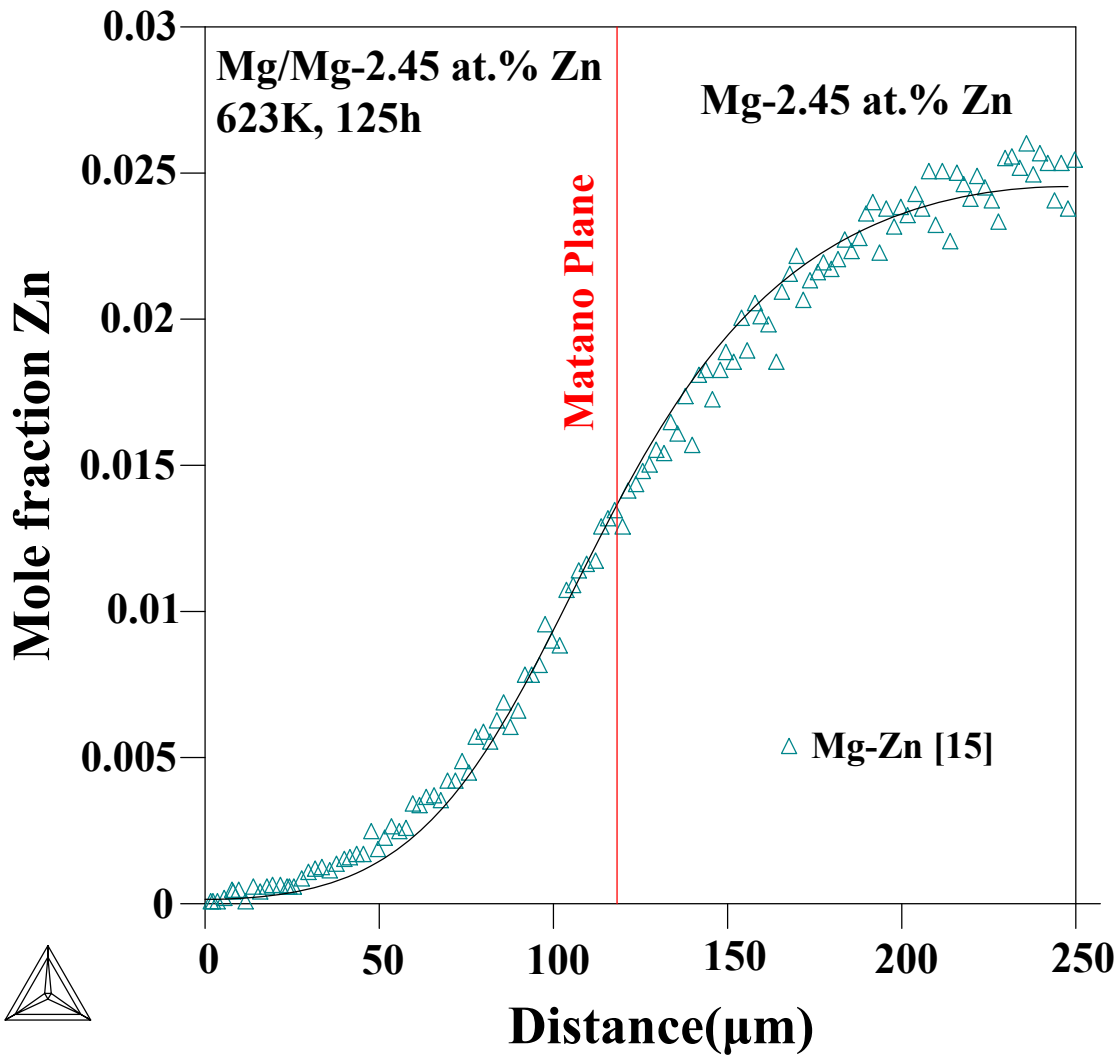

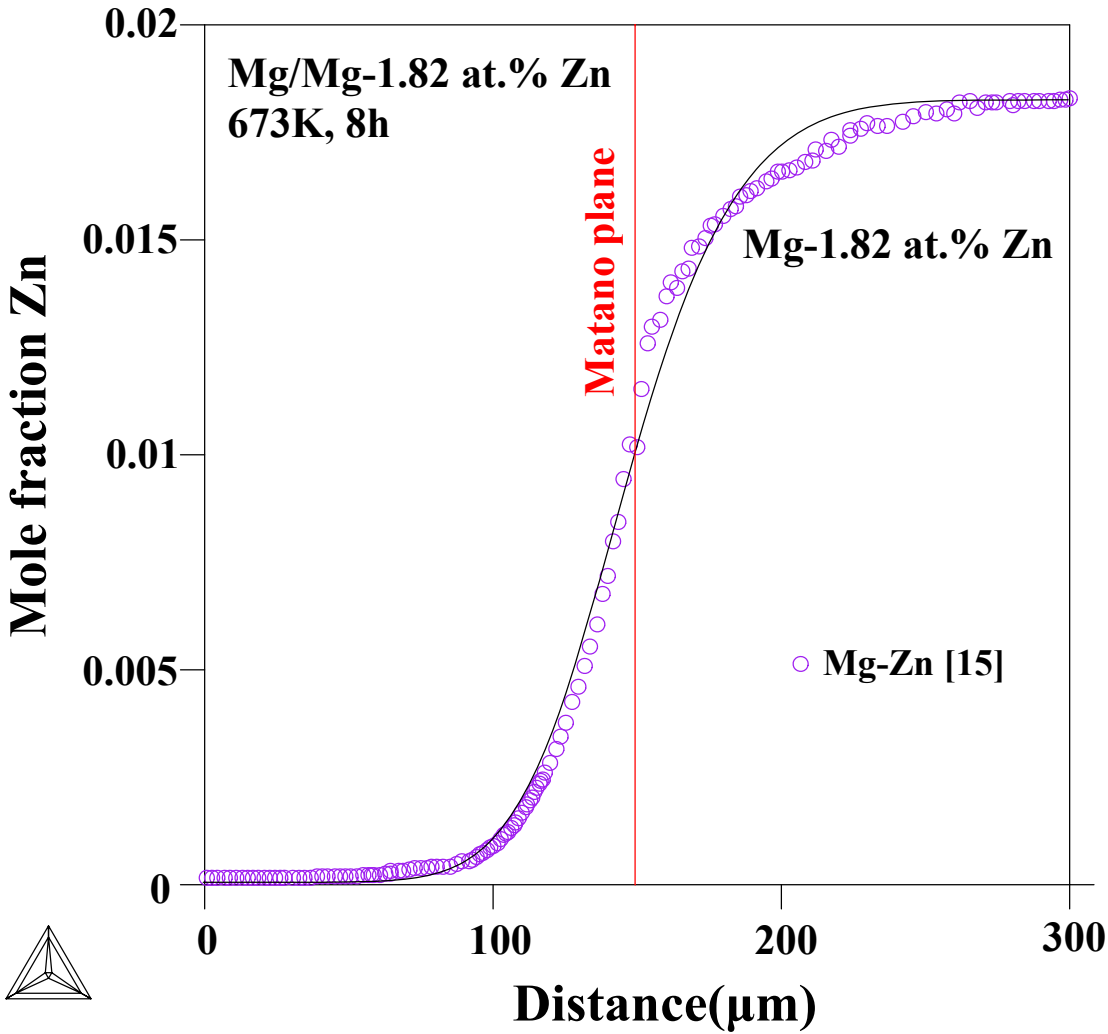

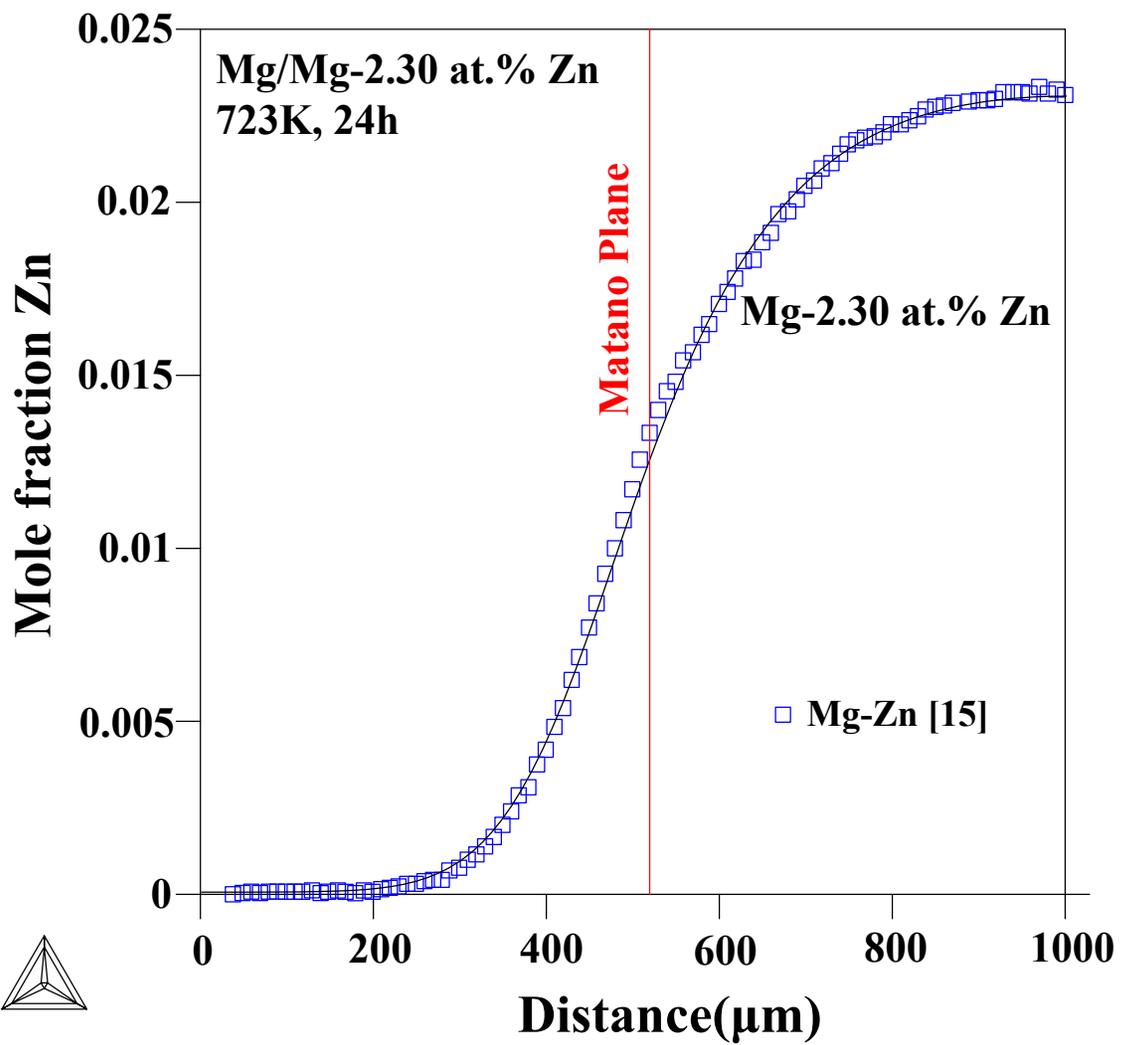

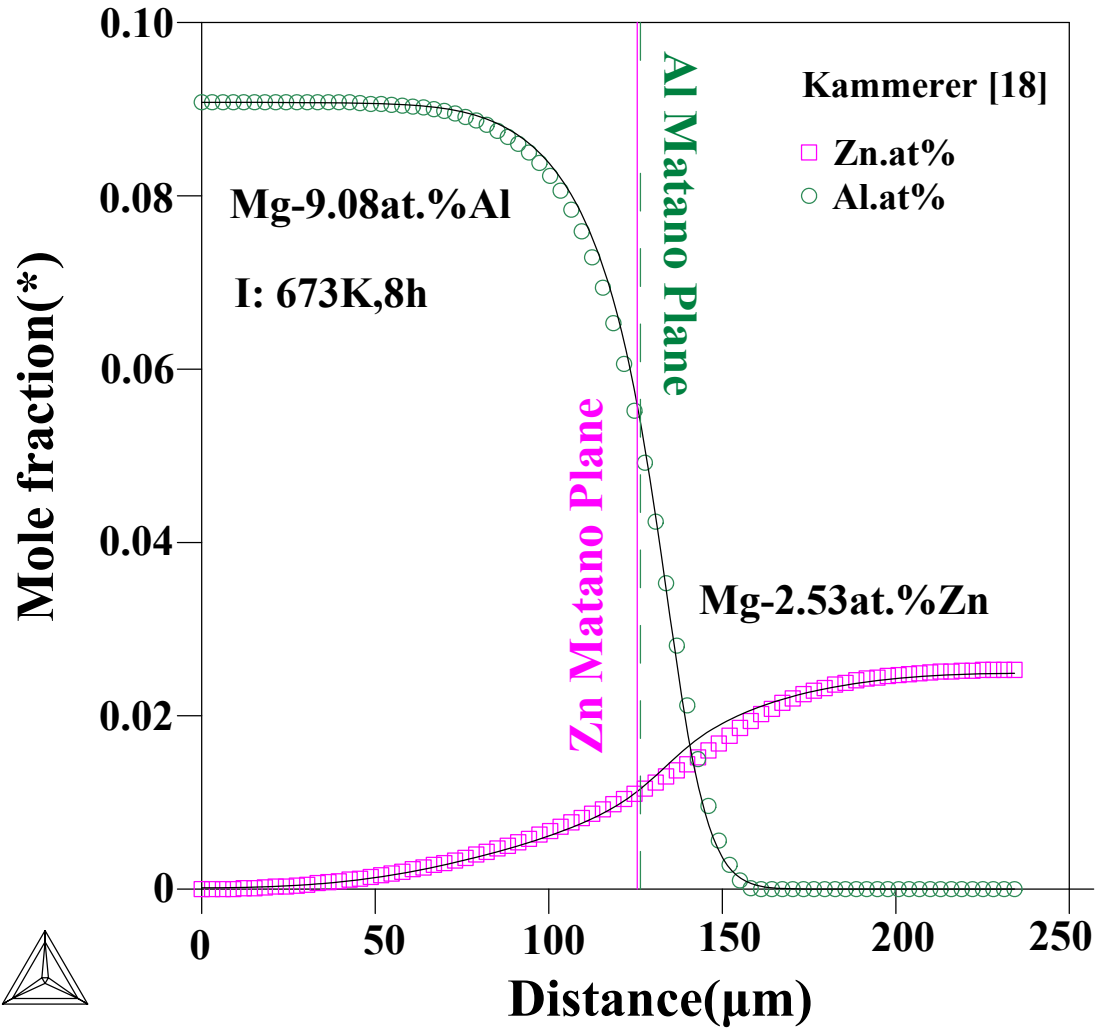

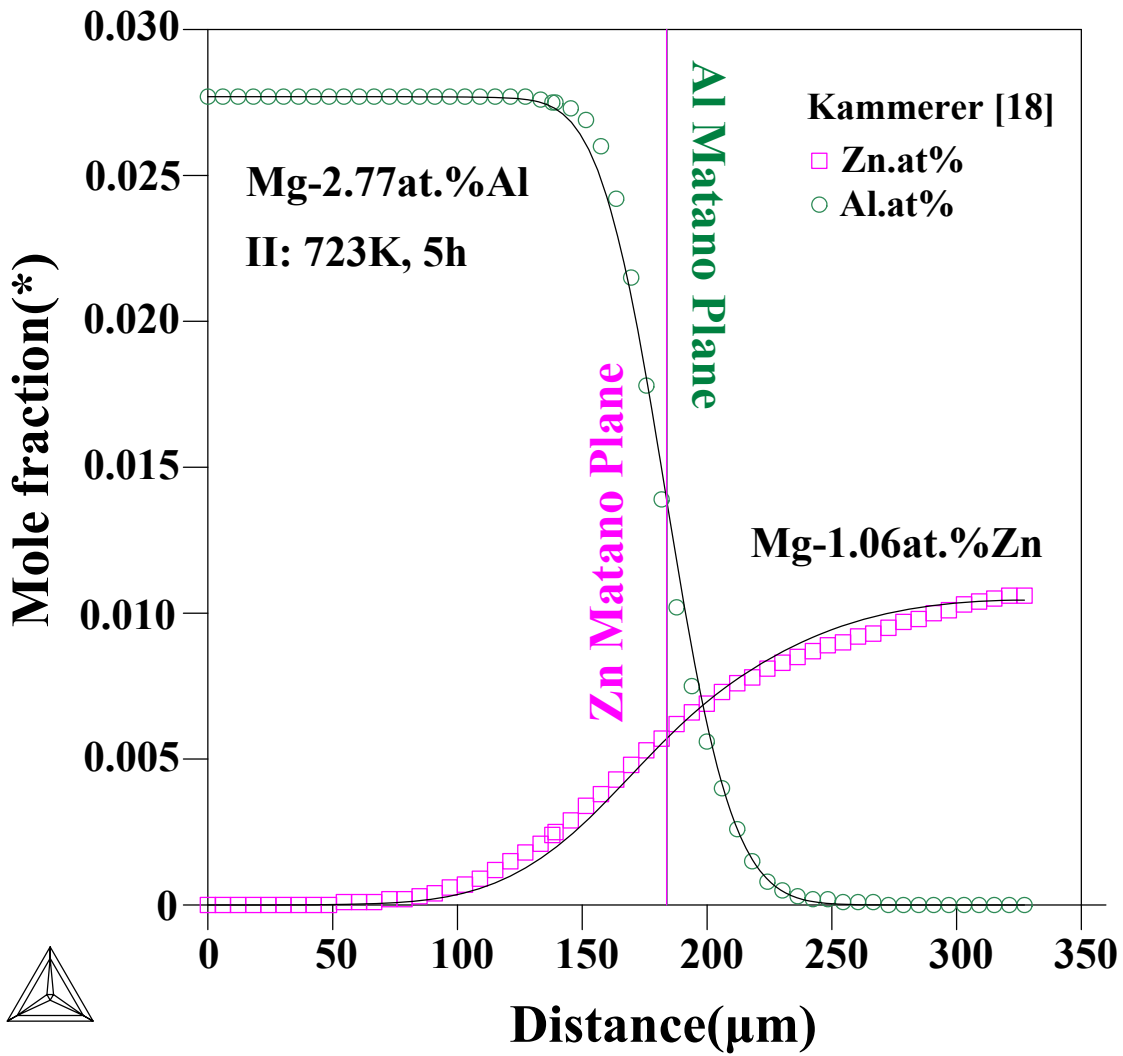

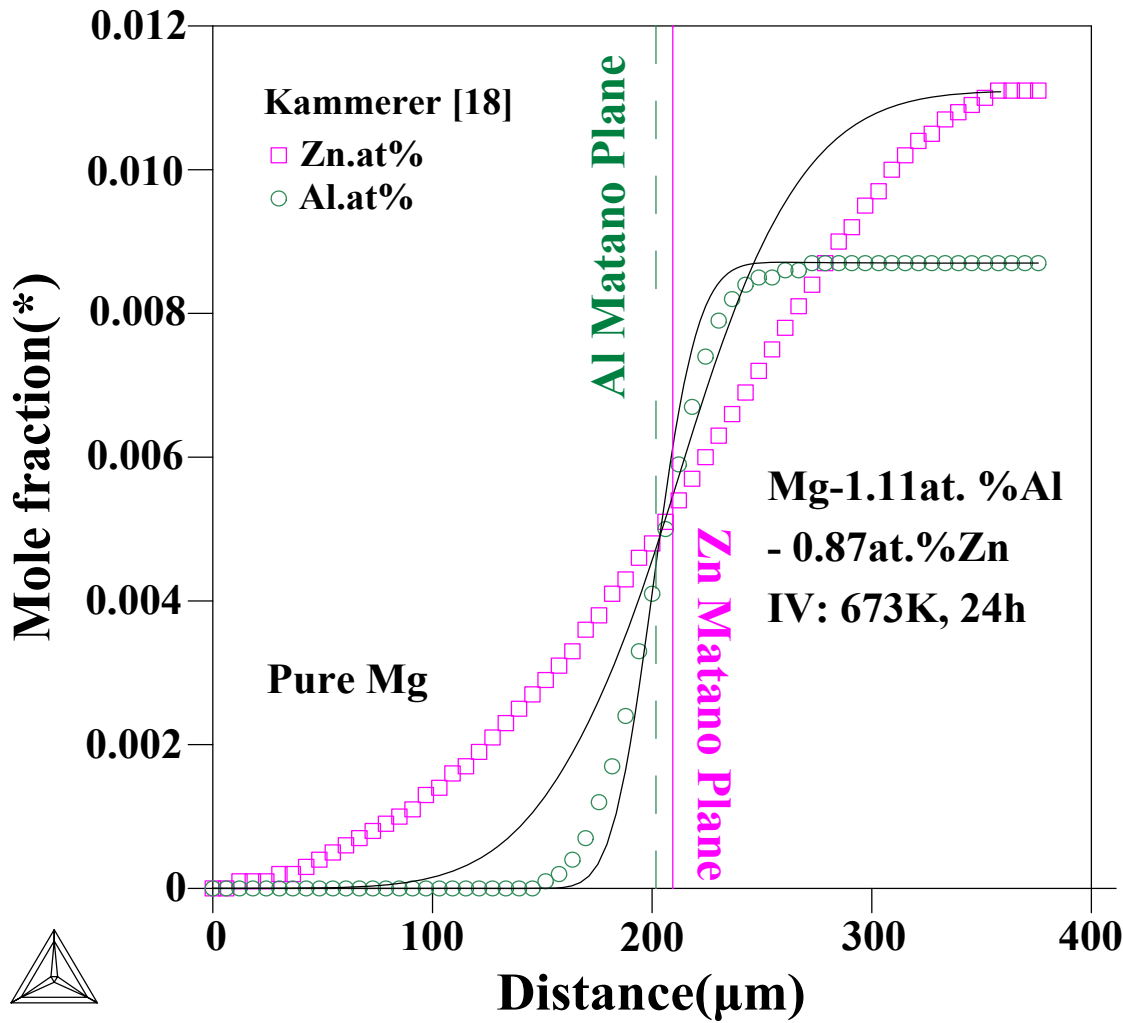

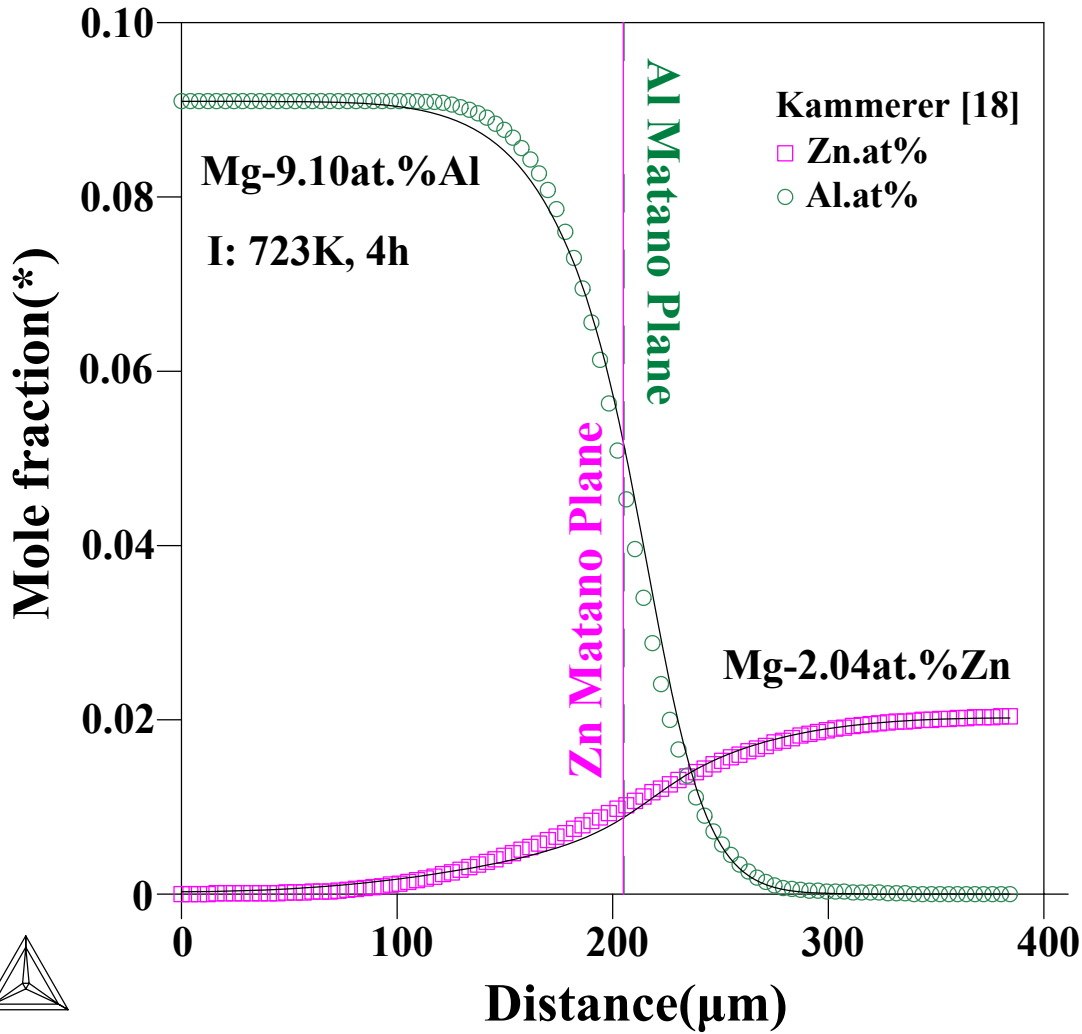

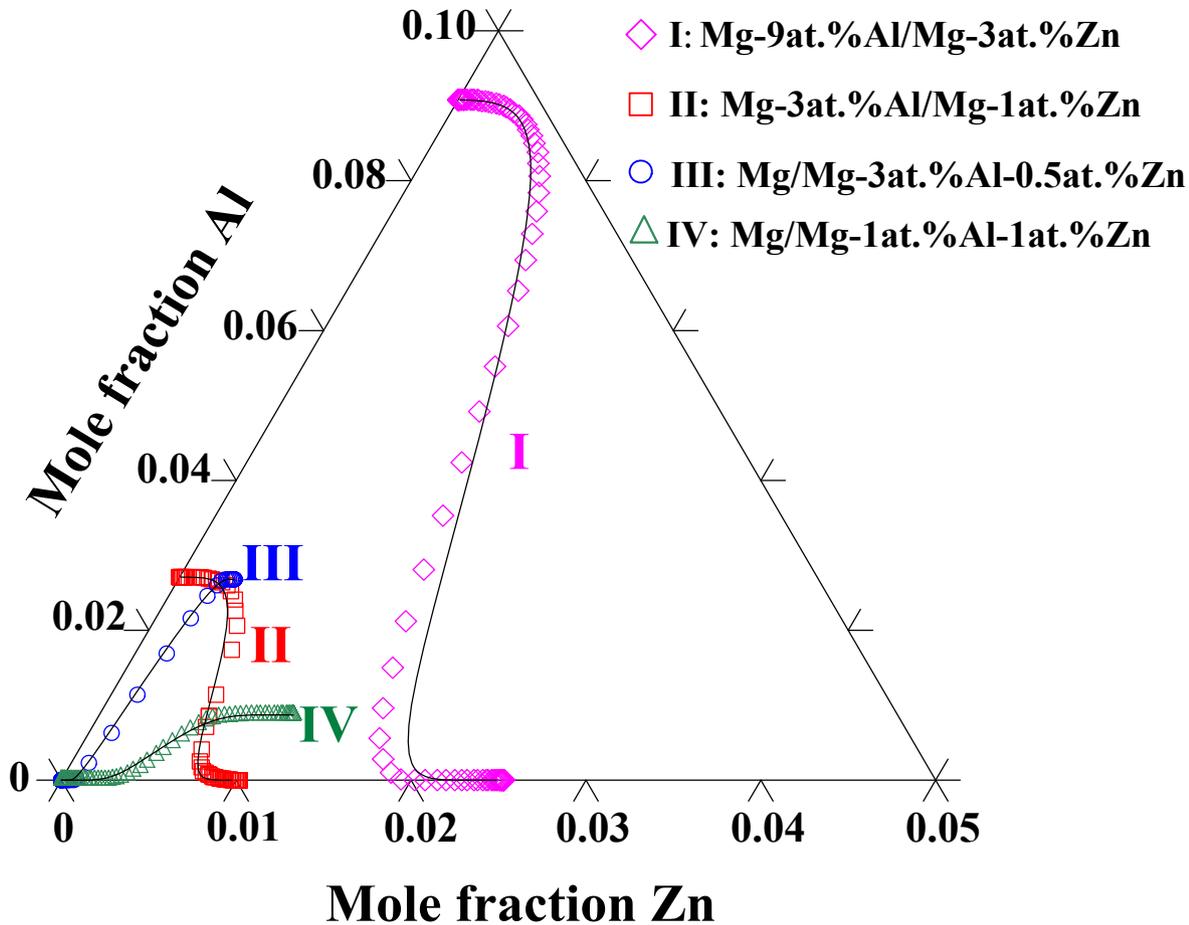

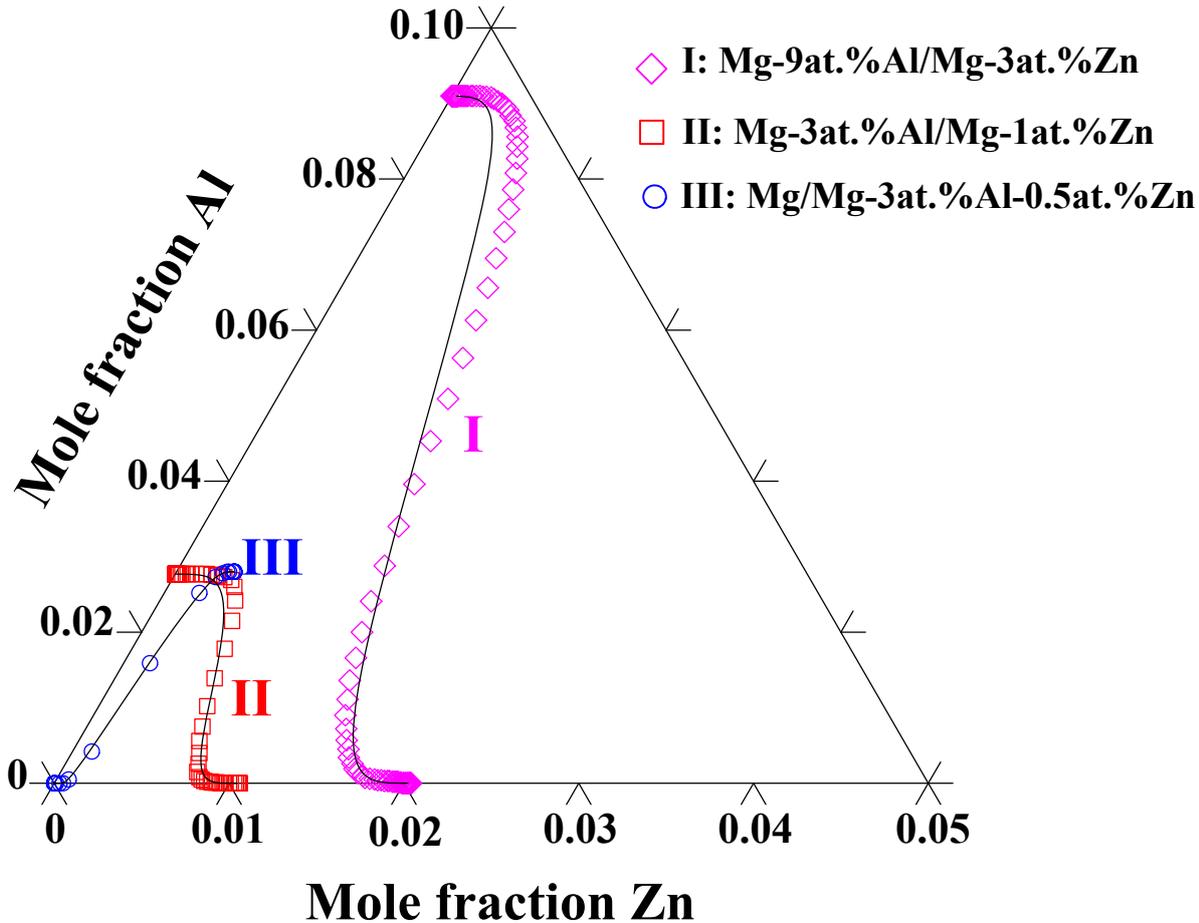